# On Tailoring Structural and Optoelectronic Properties of TiO$_2$ Thin Films Synthesized via 'Room' Temperature High Power Impulse Magnetron Sputtering (HiPIMS)

Aarati Chacko,[1] Erwin Hack,[2] Sebastian Lohde,[2] Robin Bucher,[2] Oguz Yildirim,[1] Arnold Mueller,[3] Michel Calame,[2,4,5] Hans J. Hug,[1,5] Mirjana Dimitrievska[2*]

*1 – Functional and Magnetic Thin Films, Swiss Federal Laboratories for Material Science and Technology (EMPA) Ueberlandstrasse 129, 8600 Duebendorf, Switzerland*

*2 - Transport at Nanoscale Interfaces Laboratory, Swiss Federal Laboratories for Material Science and Technology (EMPA) Ueberlandstrasse 129, 8600 Duebendorf, Switzerland*

*3 - Laboratory of Ion Beam Physics, ETH Zurich, CH-8093 Zurich, Switzerland*

*4 – Swiss Nanoscience Institute, University of Basel, 4056 Basel, Switzerland*

*5 – Department of Physics, University of Basel, CH-4056, Basel, Switzerland*

*\*corresponding author:* mirjana.dimitrievska@empa.ch

**Titanium dioxide (TiO$_2$) is a key material in optoelectronic and energy conversion technologies, including solar cells and photocatalysis. However, integrating TiO$_2$ into flexible or temperature-sensitive devices requires deposition techniques that avoid high-temperature processing while maintaining control over both phase composition and crystallinity. In this work, we demonstrate the synthesis of nanocrystalline TiO$_2$ thin films using High Power Impulse Magnetron Sputtering (HiPIMS) at near-room temperature. By systematically varying total pressure and oxygen flow, we achieve tunable anatase-to-rutile phase ratios and control over crystalline quality, as evidenced by Raman and photoluminescence trends. The observed optical trends—in both refractive index and emission—are directly linked to the underlying structural evolution, with compositional analysis verifying stoichiometric consistency across all deposition conditions. Our findings establish HiPIMS as a powerful low-temperature method for tailoring TiO$_2$ thin films and enabling their application in flexible photovoltaics, photoelectrochemical water splitting, and other energy-related systems.**

**Introduction**

Titanium dioxide (TiO$_2$, titania) is a material of wide-ranging technological importance, across fields that include optoelectronics [1–4], energy conversion [5–8], and biomedical devices [9–11]. Among these, TiO$_2$ has become especially prominent in the field of energy technologies due to its chemical stability, wide bandgap, and favorable electronic properties. It serves a critical function in photovoltaics, particularly as an electron transport layer in dye-sensitized and perovskite solar cells [12,13], and is also widely used in photoelectrochemical (PEC) water splitting systems [14,15], where its photocatalytic activity under ultraviolet and visible light enables efficient solar-to-chemical energy conversion.

While TiO$_2$ can be synthesized in many crystal structures [16], rutile and anatase are the most commonly studied, especially in thin film form. Magnetron sputtering (MS) is an inherently energetic process as the coating flux must transfer energy to the substrate when reflecting-off or condensing-onto the growing film. This is important for compounds like titanium dioxide: unless provided sufficient energy, the TiO$_6$ octahedral building blocks of TiO$_2$ will not take up crystalline arrangements, growing instead as amorphous films. Sputter-coating methods, such as direct current (DCMS) and radiofrequency (RFMS) magnetron sputtering, do not intrinsically reach this energy requirement. Instead additional energy is provided to crystallize the films through extrinsic methods that include substrate heating or post-deposition annealing typically to temperatures above 400 °C [17]. This makes depositing crystalline TiO$_2$ on temperature-sensitive substrates a challenge.

The demand for low-temperature deposition techniques is rapidly growing, particularly as energy conversion devices evolve toward lightweight, flexible, and wearable formats. Substrates such as polymers, foils, and composite materials — commonly used in flexible solar cells [18,19], roll-to-roll printed photodetectors [20], and portable water-splitting systems [21,22] — cannot withstand conventional high-temperature processing without degradation. For instance, substrates like polyethylene terephthalate (PET), polyimide, or polyetheretherketone (PEEK) often exhibit thermal stability limits below 200 °C [23], making traditional TiO$_2$ synthesis routes incompatible. Achieving nanocrystalline or phase-controlled TiO$_2$ at such low temperatures is essential to maintain the desired electronic, optical, and catalytic properties while ensuring device integrity.

In this context, High Power Impulse Magnetron Sputtering (HiPIMS) emerges as a promising technique to produce crystalline TiO$_2$ 'at room temperature', or more accurately, *without providing any additional heat* to the film than is generated by the sputter process. Unlike conventional DCMS and RFMS, HiPIMS generates a dense and energetic coating flux, with a high degree of ionization of film-forming species. Not only does this technique inherently meet the energy requirements for crystalline TiO$_2$ [24], but HiPIMS also provides additional control mechanisms to fine tune crystal structure.

Control over the polymorphic composition of TiO$_2$ films is crucial for tailoring its properties. Rutile, generally considered the stable form of TiO$_2$, requires more energy to form from an amorphous state than the other, more metastable crystal polymorphs [17]. Anatase, once formed, can rearrange into rutile under conditions that provide sufficient additional energy. HiPIMS shows potential for growing both polymorphs [25–30], with rutile forming preferentially in environments that increase the energy of film-forming species, such as high peak power density [26,28,29], biasing [25,30] and low deposition pressure [16,24,26], as well as substrate heating [31,32]. In contrast, anatase preferentially grows under weaker ion bombardment [24,26]. Low temperature HiPIMS $TiO_2$ films are generally nanocrystalline and often polymorphic. The specific mix and structure of polymorphs, and resulting film properties, can be controlled by adjusting deposition parameters, and the literature points mainly to deposition pressure and pulse frequency [16,25,26,30,33] (which strongly influences peak power density) for low-temperature control levers. Although these studies outline HiPIMS's capabilities for growing nanocrystalline titania at low temperature, there is still a relative paucity of detailed and systematic synthesis-structure-property-relation studies for HiPIMS $TiO_2$.

In this work, we systematically investigate the influence of HiPIMS synthesis parameters on the structural and optoelectronic properties of $TiO_2$ films grown in a mid-sized, high vacuum chamber at near-ambient temperatures. Specifically, we probe the effect of total deposition pressure and oxygen partial pressure (approximated by relative oxygen flow rate) on the resulting film properties. These parameters span an as yet under-explored parameter space (see **Figure 1**). Morphological characterization was obtained through atomic force microscopy (AFM) and electron microscopy (EM). Atomic composition was determined using Rutherford Backscattering Spectrometry (RBS) and Heavy Ion Elastic Recoil Detection Analysis (HI-ERDA). Crystal polymorphs were identified by grazing incidence x-ray diffraction (GID) coupled with Raman spectroscopy. Finally, optoelectronic properties were assessed using photoluminescence (PL) and ellipsometry. Based on these results, we discuss the complex interplay of the pressure- and the oxygen-environment parameters, highlighting the competing mechanisms and their effects on the properties of HiPIMS $TiO_2$ thin films.

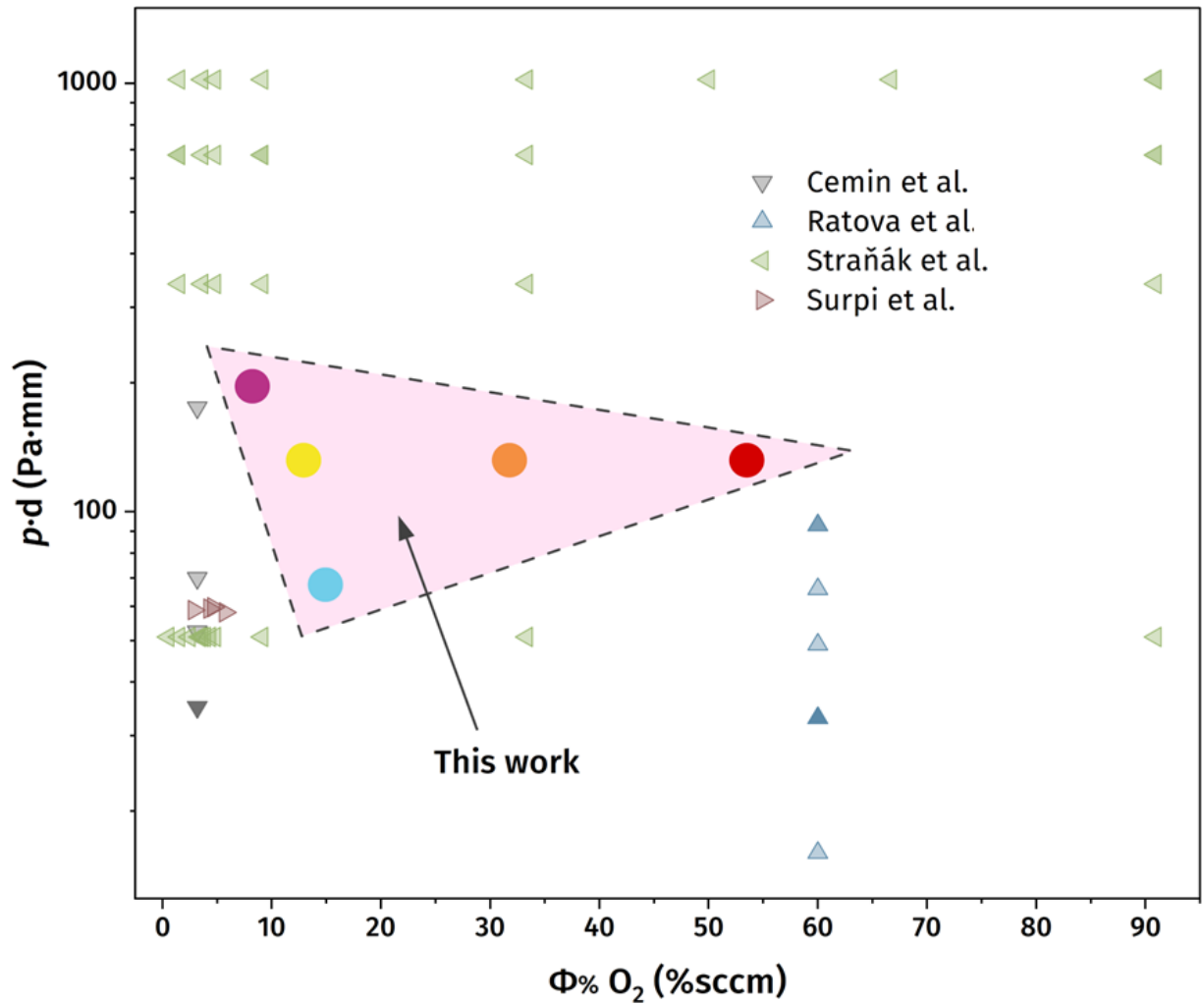

**Figure 1. Parameter space explored in this work**, compared along pressure-distance ($p$-d, in Pa·mm) and proportion of oxygen flow to the chamber ($\Phi_\%$ $O_2$, in %sccm) axes to previous studies [26,32,34,35]. The pressure-distance parameter is used as it captures coating energetics better than pressure alone.

**Materials and methods**

The $TiO_2$ films were deposited over an adhesion-promoting titanium (Ti) interlayer, creating a sample structure: substrate/Ti/$TiO_2$, with film thicknesses of 49.6 ± 1 nm for the interlayer and 128.5 ± 11 nm for the $TiO_2$ thin film. These layers were deposited in sequence — *without breaking vacuum* — in a cylindrical reaction chamber (Ø 390 × 540 mm), equipped with four radially mounted magnetrons in closed field configuration; broadly as described by Barker et al. [36], and shown in **Figure 2**. A single rectangular titanium sputter target (110 × 326 × 12 mm and 99.995% pure) was used for both the (metallic) Ti interlayer and (compound) $TiO_2$ films. The target was operated voltage-control mode using a MELEC SIPP2000USB Dual for HiPIMS pulsing, driven by an ADL GX 50/1000 DC power supply. Both the active magnetron and the substrate mount were grounded. Argon (Ar) gas was introduced at the Ti target, while oxygen ($O_2$) was distributed mid-way between the target and substrates as shown in **Figure 2a-b**. An MKS 647C mass flow controller was used, which allowed increments of 0.1 standard cubic centimeters (sccm) for Ar and 0.01 sccm for $O_2$. Chamber deposition pressure was monitored using an Inficon PSG552 gauge in increments of 0.01 Pa (0.1 μbar), and this threshold pressure was reliably sensitive to 0.1 sccm changes in Ar mass flow. Chamber temperature was assessed using a k-type thermocouple positioned a short distance above the sample mount, as can be seen in **Figure 2a-b**. Though not identical, the read-out provides an estimate of the average temperature experienced by the samples during film synthesis. Polyetheretherketone (PEEK) foil substrates (125 μm thick), intended for a separate study, were also mounted on the sample holder (**Figure 2c**). These PEEK foils served as a rough internal temperature reference; as the foils showed no obvious signs of warping or deformation after coating, the chamber temperature is unlikely to have significantly exceeded 143 °C (the polymer's glass transition temperature).

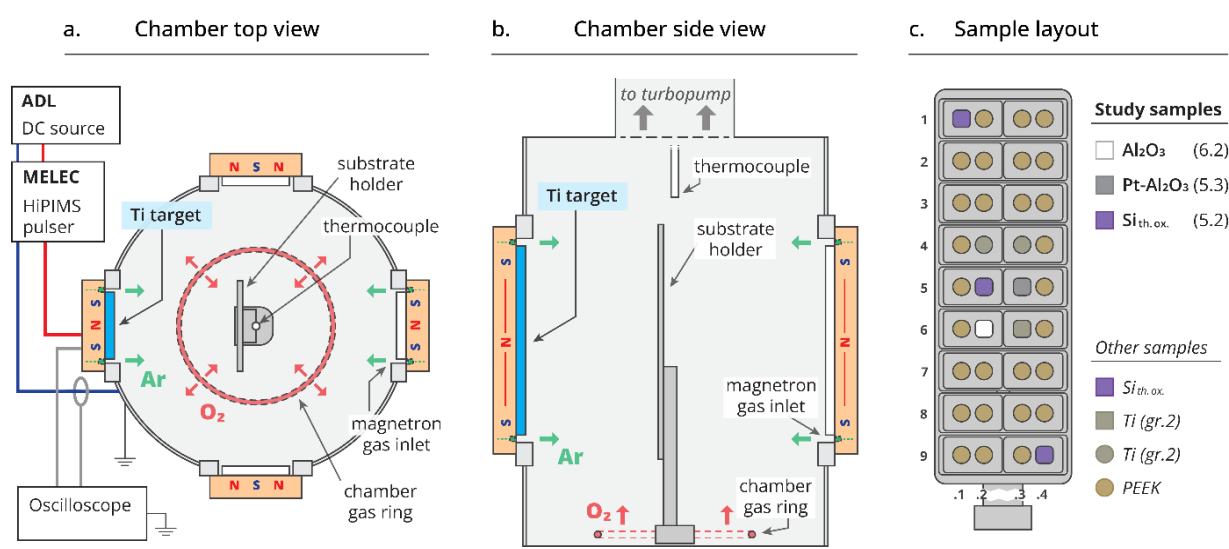

**Figure 2. HiPIMS reactor setup.** (a) Top view showing the magnetron arrangement, gas inlet points (Ar: green arrows, $O_2$: red arrows), and pulser connections. (b) Side view of the chamber with gas flow directions. The back of the substrate holder was shielded with aluminum foil (not shown). (c) Sample holder layout, showing the location of the study samples (rows 5-6, columns 2-3) and other samples.

A range of substrates were used, including thermally oxidized silicon ($Si_{th.ox.}$), (uncoated) sapphire ($Al_2O_3$), platinum-coated sapphire (Pt-$Al_2O_3$) as well as PEEK foils. To minimize the effects of sample position, substrates were assigned to type-specific locations on the sample plate (**Figure 2c**), with the study samples located centrally and adjacent to one another. Deposition parameters for the Ti interlayer were held constant for all samples: an Ar growth environment at a total pressure ($p_{Total}$) of 0.58 Pa (0.1 μbar), 592 V DC set voltage, 20 μs HiPIMS pulse length and 1000 Hz repetition frequency (2% duty cycle); with an estimated deposition rate of 8.8 nm/minute. The HiPIMS pulse parameters for the $TiO_2$ layers were also held constant: 560 V DC set voltage, 50 μs pulse length and 500 Hz pulse frequency (5% duty cycle); while chamber $p_{Total}$ and relative oxygen flow rate ($\Phi_{\%}\,O_2$) were independently varied. These measures aimed to isolate the effects that $p_{Total}$ and $\Phi_{\%}\,O_2$ have on film properties.

The key synthesis parameters are summarized, in **Table 1**, for each of the varied parameters, as well as for the Ti interlayer (please check the section Parameters selection for HiPIMS TiO$_2$ in Results and Discussion for more details on how these conditions were chosen). For samples along the $p_{Total}$ axis, we provided the lowest $\Phi_{\%}$ O$_2$ at which a stable deposition could be sustained such as to ensure similar degrees of target oxidation (poisoning) across the different pressure conditions. This was 15%, 13% and 8% for the each of the 0.21, 0.41 and 0.61 Pa $p_{Total}$ pressures. Along the $\Phi_{\%}$ O$_2$ parameter axis, we kept the $p_{Total}$ constant at 0.41 Pa, varying instead the $\Phi_{\%}$ O$_2$ (and consequently, the degree of target poisoning). For this, two additional $\Phi_{\%}$ O$_2$ values were established: 32.4% and 53.3 $\Phi_{\%}$ O$_2$, providing increasingly oxygen-rich synthesis environments. In total, five different TiO$_2$ thin films were grown, probing three $p_{Total}$ conditions and three oxygen $\Phi_{\%}$ O$_2$ conditions. To ensure reproducibility of the synthesis, each deposition was performed twice on each substrate type, with no significant variations observed between runs.

**Table 1.** Summary of synthesis parameters for HiPIMS Ti and TiO$_2$ layers. Controlled parameters, $\Phi_{\%}$ O$_2$ and $p_{Total}$ are highlighted. The other reported parameters were measured rather than controlled. These are: peak current density (I$_{peak}$), estimated average temperature (T$_{av}$), the deposition time and film growth rate.

| Parameters → <br> ↓Layer/Set | $\Phi_{\%}$ O$_2$ <br> (%sccm) | $p_{Total}$ <br> (Pa) | I$_{peak}$ * <br> (mA/cm$^2$) | T$_{av}$ ** <br> (°C) | Deposition time <br> (minutes) | Deposition rate <br> (nm/minute) |
|---|---|---|---|---|---|---|
| Ti interlayer | 0.00 | 0.58 | 266 | 49.8 | 5.6 | 8.80 |
| TiO$_2$ set: *Increasing* $\Phi_{\%}$ O$_2$ | 12.90 | 0.41 | 253 | 81.6 | 62.8 | 2.10 |
|  | 31.75 | 0.41 | 311 | 86.2 | 76.7 | 1.49 |
|  | 53.49 | 0.41 | 377 | 93.2 | 104.6 | 1.35 |
| TiO$_2$ set: *Increasing* $p_{Total}$ | 14.90 | 0.21 | 263 | 76.7 | 71.8 | 1.76 |
|  | 12.90 | 0.41 | 253 | 81.6 | 62.8 | 2.10 |
|  | 8.21 | 0.61 | 256 | 86.8 | 76.3 | 1.50 |

Notes:  * I$_{peak}$ was calculated using the exposed (unshielded) target area (283.36 cm$^2$).
   ** T$_{av}$ for the TiO$_2$ layers was estimated using the equilibrium temperature (see *Sample synthesis*).

**Characterization**

**Substrate selection.** To ensure accurate characterization and avoid substrate-induced artifacts, all films were deposited on optically and structurally appropriate substrates tailored to each measurement technique. Raman and photoluminescence (PL) measurements were performed on TiO$_2$ films grown on single-side polished, uncoated sapphire (Al$_2$O$_3$) substrates, chosen for their excellent optical transparency and minimal background signal. Atomic force microscopy (AFM) was conducted on samples grown on thermally oxidized silicon wafers to ensure consistent contrast and reliable height calibration. Scanning and transmission electron microscopy (SEM/TEM) were carried out on films deposited on silicon substrates, which provided the necessary electron transparency and conductivity. Ellipsometry measurements were performed on TiO$_2$ films grown on platinum-coated sapphire substrates (Pt/Al$_2$O$_3$) to avoid signal distortion from the uncoated Al$_2$O$_3$ interface and to ensure accurate optical modeling. RBS and HI-ERDA analyses were also conducted on sapphire-based samples due to their minimal elemental background and thermal stability. Grazing incidence X-ray diffraction (GID) was performed on sapphire substrates to take advantage of their flatness and crystallographic compatibility. Additionally, PEEK (polyether ether ketone) substrates were used in selected deposition trials (not shown in detail here) to demonstrate compatibility with temperature-sensitive applications. The absence of any visible damage or deformation of the PEEK films after deposition confirmed that the entire process remained below critical thermal thresholds (<200 °C), thereby validating the low-temperature character of the synthesis method.

**Raman and photoluminescence (PL)** spectra were collected using a WITec Alpha 300 R confocal Raman microscope operated in a backscattering configuration. Excitation wavelengths of 532 nm (for Raman) and 488 nm (for PL) were employed. A microscope objective was used to focus the laser beam onto the sample surface, yielding a spot size of approximately 1 μm, with incident laser power around 0.3 mW. Prior to measurement, a laser power-dependent stability test was performed at a fixed sample location. This test involved gradually increasing the laser power from the minimum setting while observing any spectral changes—specifically in peak position, full width at half maximum (FWHM), and the emergence of additional features. The maximum laser power that did not induce spectral changes was selected as the optimal power for all subsequent measurements. The backscattered signal was analyzed using a 300 mm focal length lens-based spectrometer equipped with diffraction gratings of 150 g/mm for PL and 1800 g/mm for Raman, along with a thermoelectrically cooled CCD detector. Raman shift calibration was carried out using the characteristic Si peak at 520 cm$^{-1}$ as an internal standard.

**Atomic Force Microscopy (AFM)** was performed on a Bruker Dimension Icon 3 microscope, using the NanoScope acquisition software (version 9.7). The samples, grown over thermally oxidized silicon substrates, were imaged in PeakForce Tapping mode, with Bruker ScanAsyst-Air cantilevers. Gwyddion software (version 2.65) [1] was used for image processing and roughness measurements.

**Scanning/Transmission Electron Microscopy (SEM/TEM)**: Focused ion beam (FIB) cross-sections of the thin films were prepared in a Helios 660 instrument for the $\Phi_{\%}$ O$_2$ samples grown on the silicon substrates, and these were subsequently imaged with a Jeol JEM2200FS transmission electron microscope (TEM). Topographical imaging of all samples was obtained using a Hitachi S-4800 scanning electron microscope (SEM) in secondary electron mode. The accelerating voltage was set to 2 keV and the emission current ranged between 1500 and 1900 nA.

**Ellipsometry:** Measurements of film thickness and refractive index were performed using a spectroscopic ellipsometer (M-2000 VI, J. A. Woollam Co., Inc., Lincoln, United States) in the wavelength range 370–1690 nm with an integration time of 5 s; on samples grown over the Pt-Al$_2$O$_3$ substrates, to preclude confounding contributions from the uncoated Al$_2$O$_3$ substrates. Using three incidence angles of 50°, 60° and 70° allowed us to simultaneously obtain the optical parameters and the film thickness. The films were modelled using Tauc-Lorentz oscillator model, which is Kramers-Kronig consistent and widely used for amorphous and nanocrystalline semiconductors, especially near their optical band edges. Optical model parameters and data fits are presented in the Supporting information.

**Rutherford Backscattering Spectrometry (RBS)** and **Heavy Ion Elastic Recoil Detection Analysis (HI-ERDA):** Compositional measurements were performed at the Laboratory of Ion Beam Physics at ETH Zurich using RBS and heavy ion (HI-ERDA). For the RBS measurements, a 2-MeV $^4$He ion beam was used, which was measured with a Si PiN diode under a backscattering angle of 167.5°. The collected data were evaluated using the RUMP program [37]. A 13-MeV $^{127}$I ion beam was chosen for the HI-ERDA measurements, where recoil ions under a scattering angle of 36° are identified by the combination of a time-of-flight (TOF) spectrometer and gas ionization chamber (GIC) [38]. Elemental depth profiles in the mass range from H to Ti were extracted by the POTKU code [39].

**Grazing Incidence x-ray Diffraction:** GID measurements were performed at a grazing incident angle of 0.5° on a Bruker Discover Plus. The instrument was equipped with a rotating anode and a Dectris Eiger2 500K detector operating in 1D mode. Collimating optics of 300 micron were used to select the beam shape.

**Results and Discussion**

**Parameters selection for HiPIMS TiO$_2$**

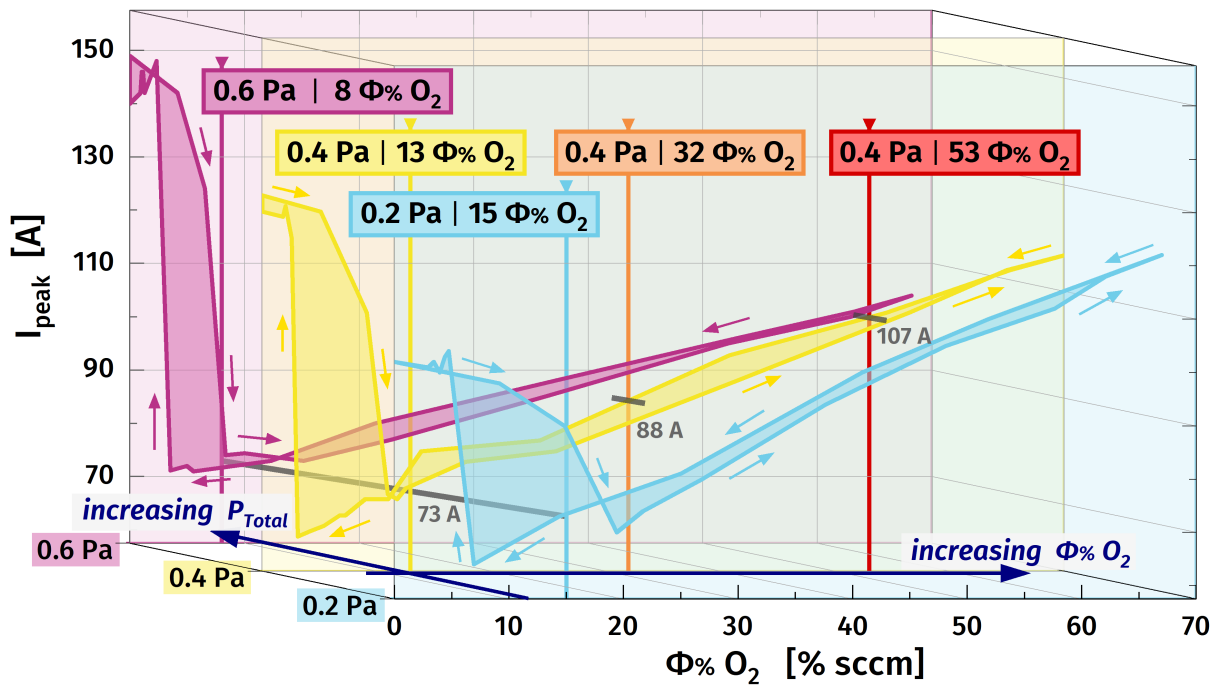

**Figure 3. Constant-pressure hysteresis curves.** HiPIMS peak current ($I_{peak}$) is plotted as a function of relative oxygen flow rate ($\Phi_{\%}$ O$_2$) for each total run pressure ($p_{Total}$) investigated in this work. Colored arrows trace the forward (metallic-to-compound) and reverse (compound-to-metallic) transitions as $\Phi_{\%}$ O$_2$ is increased and then decreased. Markers indicate the selected $p_{Total}$ and $\Phi_{\%}$ O$_2$ control parameters pairs. The color scheme identifies each synthesis condition and is used throughout this work.

The growth environments for HiPIMS TiO$_2$ were selected relative to constant-pressure hysteresis curves, shown in **Figure 3**. These curves detail the relationship between oxygen flow rate ($\Phi_{\%}$ O$_2$) and HiPIMS peak current (I$_{peak}$) at different total pressures ($p_{Total}$), and relate the Ti target surface condition to the different pressure and oxygen-flow regimes. To isolate the synthesis-property relation for $p_{Total}$, comparable target surface conditions (i.e., similar degrees of 'poisoning' or TiO$_x$ coverage) are needed across the $p_{Total}$ series. A single $\Phi_{\%}$ O$_2$ does not achieve this because the location of forward (metallic-to-compound) and reverse (compound-to-metallic) transitions vary considerably with $p_{Total}$ (**Figure 3**). For example, at 15% $\Phi_{\%}$ O$_2$, the target surface is closer to the forward transition (less poisoned) at 0.2 Pa than at 0.6 Pa. This is crucial because proximity to the transition zone significantly affects the deposition rate, and more broadly, the coating energetics. Therefore, pressure-specific $\Phi_{\%}$ O$_2$ values were chosen, at comparable distances to the forward transition points. These values were 15%, 13% and 8% $\Phi_{\%}$ O$_2$ for $p_{Total}$ pressures of 0.21, 0.41 and 0.61 Pa, respectively.

The intermediate $p_{Total}$ pressure (0.41 Pa) was chosen to build the increasing oxygen flow parameter set. Starting at 13% $\Phi_{\%}$ O$_2$, two additional values were selected: 32%, and 53%, corresponding to increasingly oxygen-rich environments at a constant $p_{Total}$ chamber pressure (0.41 Pa). The selected $p_{Total}$ and $\Phi_{\%}$ O$_2$ parameter pairs are shown (color-coded signposts) in **Figure 3**. This color scheme is used throughout this work to identify sample synthesis conditions.

**Sample synthesis**

The depositions were performed in voltage control mode at -594 V for the Ti interlayer and -560 V for the TiO$_2$ films, with constant HiPIMS pulse parameters for each (see **Figure 4a**). The key deposition stages — Ti interlayer deposition, target poisoning and TiO$_2$ thin film deposition — were carried out in sequence, without breaking vacuum. The interlayer deposition was brief (under 5.5 minutes), with chamber temperatures reaching 40-60 °C (**Figure 4b**, *Ti interlayer*). Sample-mount rotation by 180° marked the end of interlayer deposition, turning the sample surfaces away from target while exposing the foil-shielded back of the mount (cf. **Figure 2a-b**). The (argon) total pressure was then adjusted to the required synthesis $p_{Total}$, a value held constant for the remaining processing stages. In target poisoning, which immediately followed, the proportional flow of oxygen ($\Phi_{\%}$ O$_2$) was increased from 0% to about 70% $\Phi_{\%}$ O$_2$, before being decreased to the synthesis value determined in the hysteresis experiments (**Figure 4b**, *Target poisoning*). During this poisoning process, the thermocouple (now no longer shielded by the sample plate) registered a sharp rise as the target transitioned between metallic and compound states, followed by a more gradual rise and fall roughly synchronous with changes in $\Phi_{\%}$ O$_2$. HiPIMS TiO$_2$ layer deposition began when the sample mount was rotated back to face the target (as in **Figure 2a-b**). With $p_{Total}$ and $\Phi_{\%}$ O$_2$ held constant, both peak current and DC power values remained relatively stable, indicating consistent plasma conditions during film growth (**Figure 4b**, *TiO$_2$ thin film*). In contrast, the temperature sensor (now shielded) recorded an exponential cooling curve over the first 48 ± 10 minutes of layer deposition, followed by a temperature plateau. This suggests the sensor (thermocouple and casing), exposed during target poisoning, was initially hotter than the surrounding environment (chamber, sample plate and process gases); and that during TiO$_2$ deposition (once shielded), the sensor temperature then fell to equilibrium. Thus, the process temperature likely ranged between 75-95 °C (the plateau values) at the start of deposition, despite higher initial sensor readings. Therefore, the inherent heating provided by the HiPIMS process to the growing films was significantly lower than the 400 °C or more typically required for $TiO_2$ crystallization.

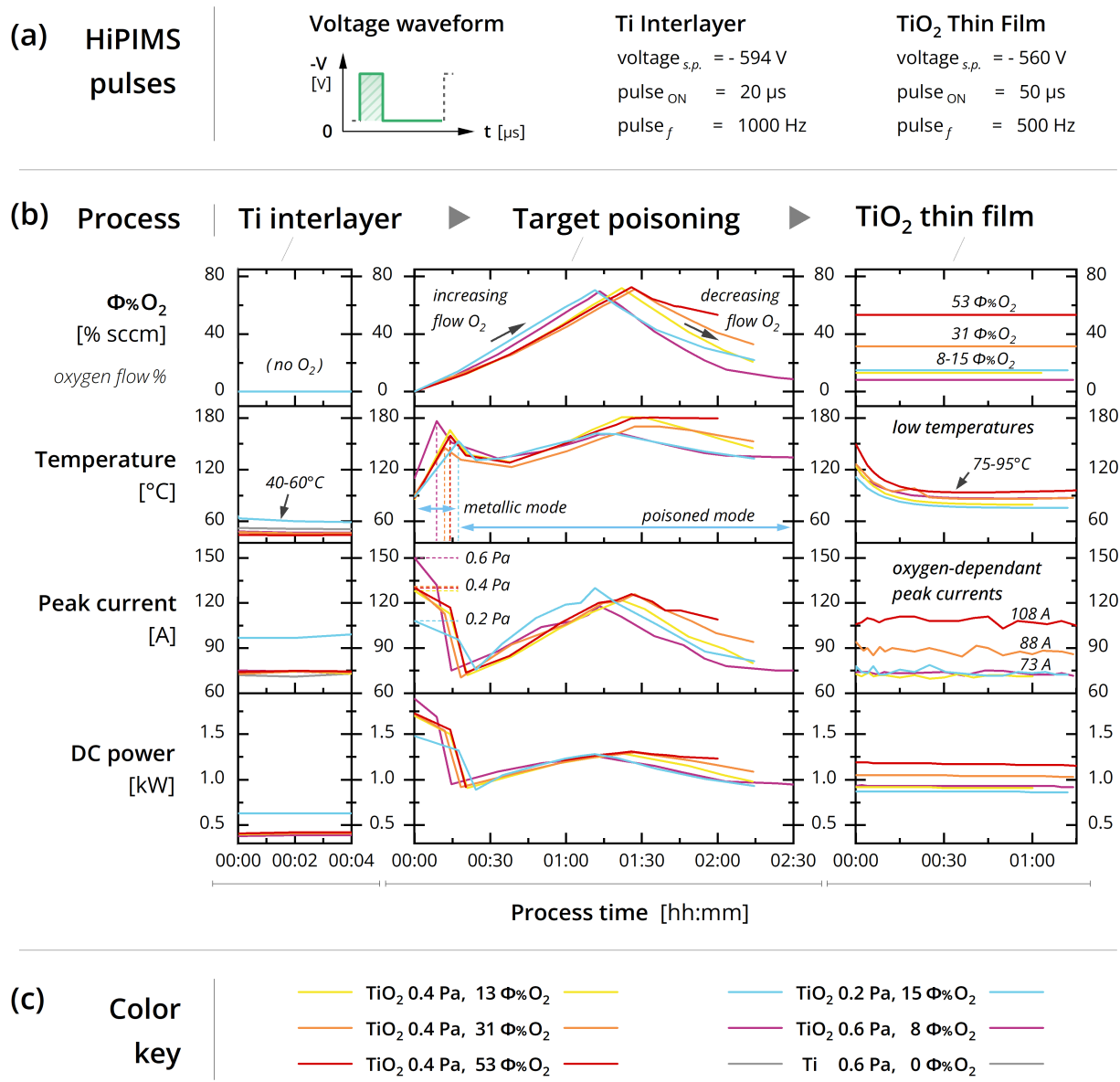

**Figure 4. Process parameters during sample synthesis. (a)** HiPIMS pulse shapes used to deposit the Ti interlayer and $TiO_2$ thin films, with constant-voltage set points. **(b)** Time evolution of process parameters: relative oxygen flow rate ($\Phi_\%\ O_2$), chamber temperature, HiPIMS peak current and DC power during the three processing stages: titanium interlayer deposition, target poisoning and $TiO_2$ film deposition. Note that temperature gains occur mainly during target poisoning, so outside of Ti or $TiO_2$ depositions. The film-forming stages themselves exhibit relatively low temperatures (40-60 °C for Ti and 75-95 °C for the $TiO_2$ films). **(c)** Color key for synthesis conditions (as in previous figures).

**Morphological and compositional assessment of HiPIMS TiO$_2$ thin films**

We begin by exploring the surface morphology and microstructure of TiO$_2$ layers synthesized under the various growth regimes. An overview of the surface morphology and cross sections for the Ti interlayer and TiO$_2$ films synthesized with increasing $\Phi_\%$ O$_2$ is shown in **Figure 5**; while the surface morphologies of films grown along the $p_{Total}$ parameter axis are shown in **Figure S1** in the Supporting Information.

Surface imaging, using SEM (**Figure 5a-d, Figure S1a-d**) and AFM (**Figure 5e-h, Figure S1e-h**), reveal significant topographic differences between the Ti interlayer surface and that of the TiO$_2$ films. The Ti interlayer exhibits a tightly packed, tooth- or needle-like structure; while the TiO$_2$ films, as a set, share a morphology that is more cauliflower-like, with a much smaller density of extreme aspect ratios than the Ti interlayers over which they were deposited. Moreover, the titania films have similar grain sizes, between 20-30 nm, irrespective of their synthesis conditions. All the synthesized TiO$_2$ layers are nanorough, with a roughness in the range of 6-8 nm as shown in **Table 2**.

Nevertheless, qualitative differences can be observed in the bulk TiO$_2$ texture of films examined using bright-field (BF) TEM cross-sectional imaging (**Figure 5i - k**). These samples, all grown at 0.4 Pa, but in increasingly oxygen-rich environments, exhibit notable differences. We see well defined columnar growth, especially in films grown at 12.9% and 53.3% $\Phi_\%$ O$_2$. The bulk texture is especially pronounced for the TiO$_2$ film grown with highest $\Phi_\%$ O$_2$, where we see large and randomly distributed crystallites (as indicated by the red arrows in **Figure 5k)**, which contrasts strongly with the far finer, texture of the 31.4% $\Phi_\%$ O$_2$ sample. All films show a preferential orientation in the growth of grains, aligned with the tooth-like structure of the Ti interlayer.

An extensive interface layer can be observed in the TEM cross-sectional images, between the deposited HiPIMS Ti and TiO$_2$ films; characterized by a distinct contrast approximately 50 nm thick, which roughly follows the sharp protrusions of the Ti crystallites. This oxide may have developed on the freshly deposited titanium interlayer during target-poisoning, when oxygen gas is first introduced to the chamber. The slow-growing, low energy oxide that forms under these conditions, would likely have a different structure compared with TiO$_2$ synthesized through sputter processes [29]. However, it is unclear whether this native-like oxide is indeed formed during target-poisoning, as it might have been displaced in the initial stages of (the highly energetic) HiPIMS TiO$_2$ deposition. An alternative (and more likely) origin for the interfacial layer lies in the mismatch between the Ti and TiO$_2$ crystal structures. In order to accommodate the change from the hexagonal structure of Ti to the tetragonal structure of TiO$_2$, this interfacial layer is formed in a somewhat disordered structure, different from the bulk structure of the TiO$_2$ film.

Overall, we can conclude that the surface morphology of the HiPIMS TiO$_2$ films is largely invariant to the different pressure and oxygen environments under which they were grown, thus demonstrating the versatility of this deposition technique for the fabrication of thin oxide films with consistently homogeneous surface morphologies over a wide range of deposition parameters.

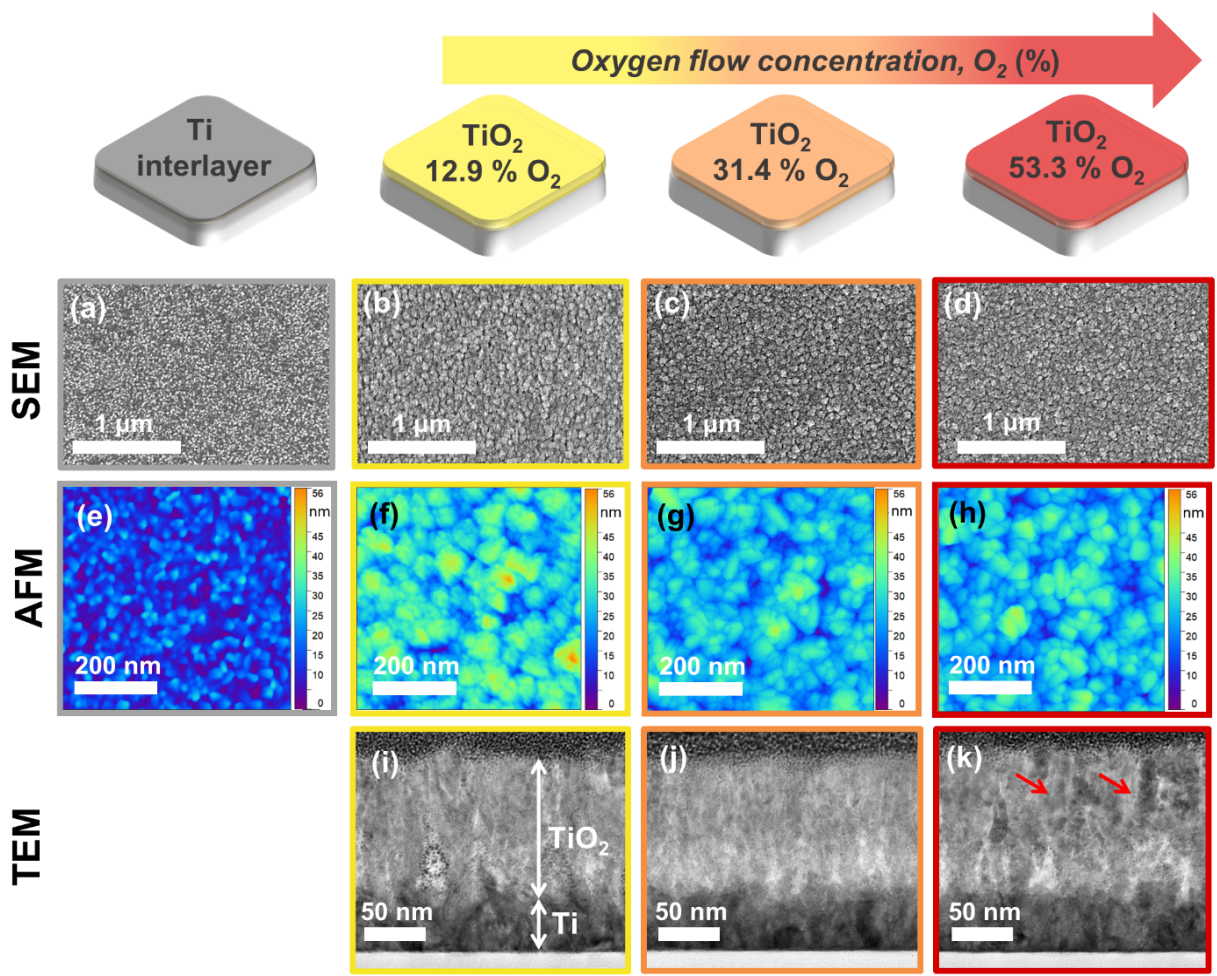

**Figure 5. Morphological assessment of Ti interlayer and TiO$_2$ thin films HiPIMS-deposited for various $\Phi$ $_\%$ O$_2$. (a) – (d)** SEM and **(e) – (h)** AFM top view images of the Ti interlayer and TiO$_2$ thin films. **(i) – (k)** Cross-sectional TEM bright field images of the TiO$_2$ thin films show changes in the microstructure upon the increase in $\Phi$ $_\%$ O$_2$.

The atomic compositions of the TiO$_2$ thin films were assessed by RBS measurements with complimentary HI-ERDA, the latter allowing for more reliable detection of light elements such as oxygen and hydrogen. The results are summarized in **Table 2**. Overall, and within the experimental error of the measurements, all TiO$_2$ thin films exhibit stoichiometric compositions, corresponding to two O atoms per Ti atom. However, it should be noted that these measurements represent the integral atomic composition across the thickness of the film, so they do not differentiate between the components (bulk vs. interfacial) or relative compositions of the crystal phases that may be present in the TiO$_2$ layer. Furthermore, minor contaminations with hydrogen and carbon are observed in all TiO$_2$ layers. These compositional values are shown in **Table S1** in the Supporting Information for TiO$_2$ films grown under the different deposition parameters.

The TiO$_2$ layer thickness were estimated using AFM and Ellipsometry, as shown in **Table 2**. The values obtained from both techniques are in good agreement, with thicknesses ranging, on average, between 120-140 nm, showing that the synthesized titania films are of comparable thickness.

**Table 2.** Compositions and thicknesses of the TiO$_2$ thin films grown under various $p_{Total}$ and $\Phi_{\%}$ O$_2$

| Film / Parameter axis | Synthesis conditions | Atomic ratio O:Ti | | Layer thickness [nm] | | Roughness [nm] |
|---|---|---|---|---|---|---|
| - | - | RBS | HI-ERDA | AFM | Ellipsometry | AFM |
| Ti interlayer | 0.58 Pa | 0.20(4) | / | 50(5) | / | 5.6(5) |
| **Increasing $\Phi_{\%}$ O$_2$** | 12.9 %O$_2$, 0.4 Pa | 2.05(8) | 2.03(15) | 133(14) | 112(12) | 7.2(5) |
| | 31.4 %O$_2$, 0.4 Pa | 2.00(10) | 2.06(15) | 115(12) | 127(13) | 6.1(5) |
| | 53.3 %O$_2$, 0.4 Pa | 1.97(8) | 2.10(15) | 142(15) | 133(14) | 6.0(5) |
| **Increasing $p_{Total}$** | 0.2 Pa, 15 %O$_2$ | 2.00(10) | 2.07(15) | 126(13) | 131(14) | 6.1(5) |
| | 0.4 Pa, 12.9 %O$_2$ | 2.05(8) | 2.10(15) | 133(14) | 115(12) | 7.2(5) |
| | 0.6 Pa, 8 %O$_2$ | 2.05(9) | 2.07(15) | 118(12) | 115(12) | 7.7(5) |

*Note :   (x) indicates the uncertainty of the measurement*

**Structural and phase assessment of TiO$_2$ thin films**

The influence of deposition conditions on the structural properties and crystal phases formed in the TiO$_2$ layers was investigated using GID and Raman measurements.

GID measurements were performed at a grazing incidence angle of 0.5°, resulting in a penetration depth of about 160 nm into the TiO$_2$ layer (**Figure 6a**). The penetration depth was calculated using the methodology presented in Ref. [40]. Considering the thicknesses of the TiO$_2$ layers (120-140 nm, as given in **Table 2**), we can therefore expect contributions from the Ti interlayer in the sample GID patterns.

An overview of the GID measurements and analysis is presented in **Figure 6**. GID patterns of the synthesized TiO$_2$ films for increasing $\Phi_{\%}$ O$_2$ and increasing $p_{Total}$, as well as the reference from substrates with Ti interlayer are shown in **Figure 6b** and **6c**, respectively. Reference patterns of the anatase (JCPDS Card no. 21-1272) [41], rutile (JCPDS Card no. 21-1276) [41], and elemental Ti (JCPDS Card no. 44–1294) [41] are reported on the x-axis below the GID patterns in **Figure 6b** and **6c**, while crystal structure illustrations of anatase and rutile TiO$_2$ are given in **Figure 6d** and **6g**, respectively. In all TiO$_2$ samples, we observe reflections at 2θ angles of 38.5° and 40.4° that originate from the Ti interlayer underneath the TiO$_2$ films that match well the reflections of the (uncoated) Ti film references (labeled as 'Ti ' in **Figure 6b** and **6c**.) LeBail refinements of patterns identify the presence of three phases, Ti (hexagonal structure, P63/mmc), anatase $TiO_2$ (tetragonal structure, $I4_1$/amd and rutile $TiO_2$ (tetragonal structure, $P4_2$/mnm). These phases are present in all the $TiO_2$ films showing that all samples are polycrystalline, as detailed next. Further characterization of $TiO_2$ films with Raman spectroscopy provide additional insights, as discussed below. Note: While brookite $TiO_2$ (orthorhombic structure, Pbca) was considered during the refinements, the measured data did not allow for a definitive conclusion about its presence or absence, suggesting either a very minor contribution or a highly disordered form of this phase.

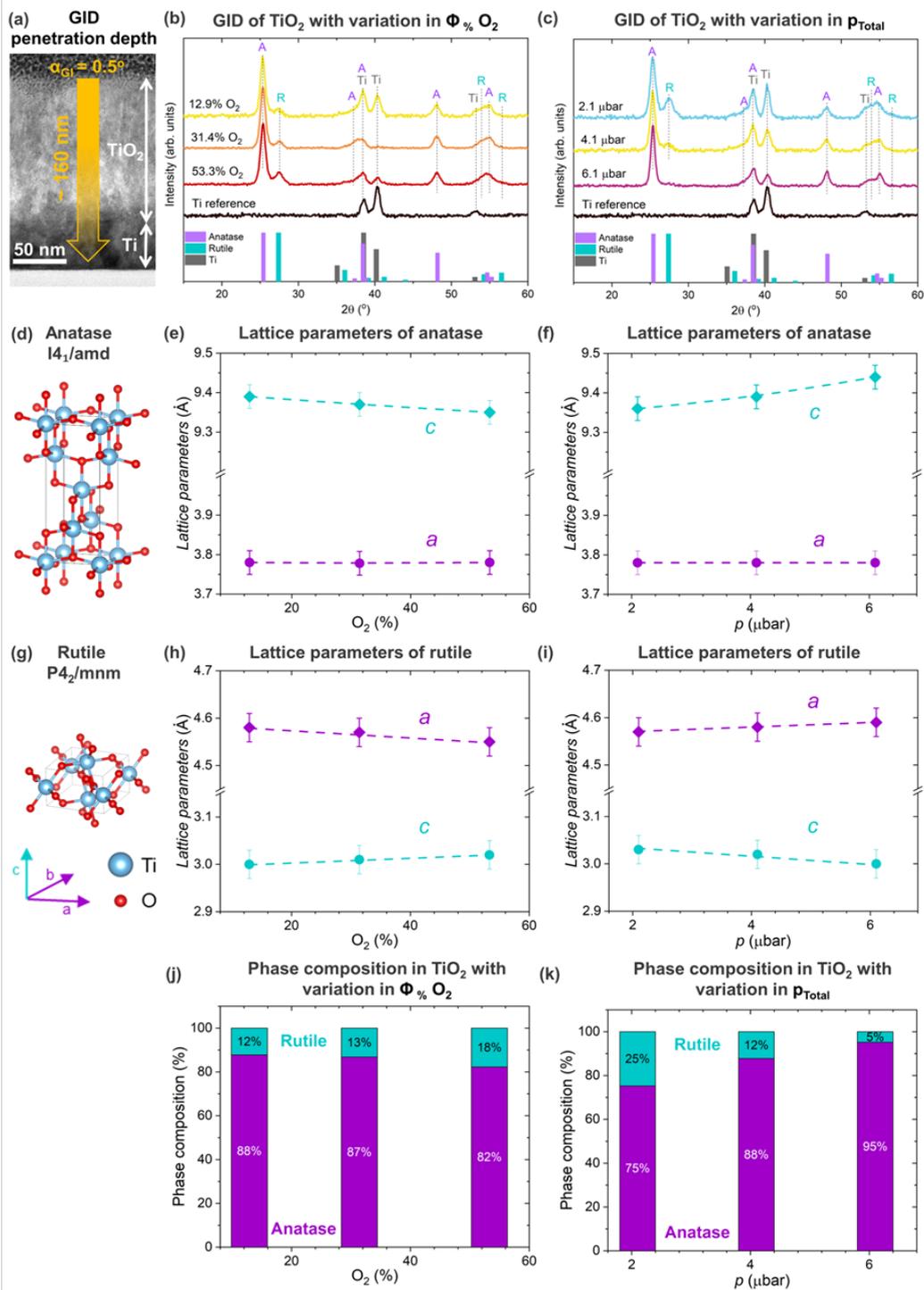

**Figure 6. Phase and structural characterization of TiO$_2$ thin films synthesized by HiPIMS: (a)** Estimated penetration depth of x-rays in TiO$_2$ thin films for GID measured at an incidence angle ($\alpha_{GI}$) of 0.5°. GID patterns of TiO$_2$ thin films synthesized under **(b)** increasing $\Phi_{\%}$ O$_2$ and **(c)** increasing $p_{Total}$. All patterns are normalized to the Bragg reflection at 25.4°, belonging to anatase phase. LeBail refinements of patterns have identified the presence of three phases, Ti (hexagonal structure, P6$_3$/mmc), anatase TiO$_2$ (tetragonal structure, I4$_1$/amd) and rutile TiO$_2$ (tetragonal structure, P4$_2$/mnm); and for which reference patterns are shown on the x-axis, below the GID patterns. Crystal structure representation of **(d)** anatase and **(g)** rutile TiO$_2$. Calculated lattice parameters of anatase TiO$_2$ phase for increasing **(e)** $\Phi_{\%}$ O$_2$ and **(f)** deposition pressure. Calculated lattice parameters of rutile TiO$_2$ phase for increasing **(h)** $\Phi_{\%}$ O$_2$ and **(i)** deposition pressure. Labels **A**, **R** and **Ti** above peaks reflect the presence of anatase phase, rutile phase and Ti interlayer, respectively. Concentration of anatase and rutile phase in TiO$_2$ films with **(j)** increasing $\Phi_\%$ O$_2$ and **(k)** increasing $p_{Total}$.

Intense Bragg reflections of anatase, located at 2θ angles of 25.4° and 48.2° correspond to the 101 and 200 crystal plane orientations. These reflections are dominant in the GID patterns of all HiPIMS TiO$_2$ films, regardless of the synthesis $\Phi_\%$ O$_2$ or deposition pressure. The presence of the rutile phase is indicated by reflections at 27.6° and 54.3°, which correspond to the 110 and 211 crystal plane orientations. An increase in the 110 peak intensity of rutile is unambiguously observed to correlate with increasing $\Phi_\%$ O$_2$ (**Figure 6b**), while a decrease in the intensity of these peaks is observed with increasing $p_{Total}$ (**Figure 6c**). These variations in peak intensity reflect changes in the quantity of the rutile phase under different synthesis parameters. While Rietveld refinement allows the most accurate determination of the phase concentration from XRD, its application was not feasible due to the nanocrystalline nature of the system. Therefore, we utilized an empirical relation that has been widely applied for determining the anatase and rutile phases in TiO$_2$ [42–44]. Relative concentrations of anatase and rutile phases in the TiO$_2$ films, for increasing $\Phi_\%$ O$_2$ and increasing $p_{Total}$, are shown in **Figure 6j** and **Figure 6k** respectively.

Increasing the $\Phi_\%$ O$_2$ at constant total pressure ($p_{Total}$) leads to a corresponding rise in the oxygen partial pressure within the deposition chamber, thereby enhancing the amount of reactive oxygen available to interact with titanium atoms during film growth. Specifically, at $p_{Total}$ = 0.41 Pa, increasing $\Phi_\%$ O$_2$ from 13% to 32.4% and 53.3% results in calculated oxygen partial pressures of approximately 0.053 Pa, 0.133 Pa, and 0.219 Pa, respectively (as detailed in the Supporting Information). These increasingly oxygen-rich environments promote the formation of the rutile phase, which is thermodynamically favored under such conditions compared to anatase [45,46]. Furthermore, the highly ionized HiPIMS plasma delivers ample energy to the adatoms, with average kinetic energies estimated to increase from approximately 0.03 eV at high pressure (0.61 Pa) to over 0.1 eV at low pressure (0.21 Pa), based on mean free path and collision frequency analysis shown in the Supporting Information. This enhanced energetic flux significantly improves surface mobility, allowing adatoms to overcome the activation barriers required for rearrangement into the more compact and thermodynamically stable rutile structure. As such, the combination of increased oxygen partial pressure and sufficient adatom mobility at lower pressures leads to a higher rutile content in the films deposited under these conditions.

Increasing the total chamber pressure at low $\Phi_\%$ O$_2$ (as per our second parameter axis) significantly reduces the mean free path of sputtered species traveling from the target to the substrate. Based on kinetic gas theory (and as detailed in Supporting Information), we estimate that the mean free path decreases from approximately 32.4 mm at 0.21 Pa to 11.2 mm at 0.61 Pa, corresponding to an increase in the number of gas-phase collisions from ~1.5 to ~4.5 over the typical 5 cm target-to-substrate distance. Assuming exponential energy attenuation with each collision, the relative kinetic energy retained by sputtered particles decreases from ~21% to just ~1% across this pressure range. This substantial energy loss leads to reduced surface mobility of adatoms, favoring the formation of the anatase phase, which is more tolerant to disorder and requires lower surface energy for nucleation and growth. In contrast, under lower pressure conditions, the higher-energy coating flux promotes enhanced adatom mobility, enabling the formation of the more compact and thermodynamically stable rutile phase. It should be noted that too much energy can be detrimental to crystal and film quality. An excessively energetic flux could dislodge (preferentially sputter away) oxygen adsorbed onto the growing film rather than help incorporate it, or even amorphize the previously formed layer [29]. To be clear, however, our GID results indicate that the coating flux energies in our processes did lie below that obliterating threshold. Instead, our films follow the general trend: we observe that increasing $p_{Total}$ leads to a change in dominant polymorph fraction, from rutile-rich films to anatase-rich films, as reported by others [26–28,47,48].

We further investigate the change in structural parameters, i.e., *a* and *c* lattice constants, obtained from LeBail refinements, of anatase (**Figure 6e** and **6f**) and rutile phase (**Figure 6h** and **6i**) for different deposition conditions. Within the experimental errors of the refinements, we do not observe changes in the lattice constants of rutile phase, which also agree well with the values reported in the literature (JCPDS Card no. 21-1276) [41]. This is expected, as the rutile phase features a tightly packed tetragonal lattice where each titanium atom is octahedrally coordinated with oxygen atoms, resulting in shorter and stronger Ti-O bonds. This dense packing and robust bonding make rutile less prone to structural deformation. Anatase, on the other hand, has a structure that is less dense and more distorted, and thus, more easily affected by the synthesis conditions.

Calculated lattice parameters of the anatase phase from LeBail refinements reveal that the *c* lattice constant is more sensitive to the changes in the deposition conditions (Figure **6e** and **6f**). This is in contrast to the *a* lattice parameter, which does not change, and is in agreement with the reported reference values in the literature (JCPDS Card no. 21-1272) [41]. With increase in $\Phi_{\%}O_2$, we observe decrease of *c* lattice parameter, which can be attributed to the incorporation of oxygen into the $TiO_2$ lattice, which can cause lattice contraction due to improved stoichiometry and a reduction in defect concentration [34,49]. In contrast, an increase in deposition pressure leads to an expansion of the *c* lattice parameter. Higher pressures result in a denser packing of the $TiO_2$ during film growth, leading to structural modifications that favor a more expanded structure. He et al. [49] noted that the self-confined growth processes at elevated pressures enhance the interlayer spacing due to increased atom mobility and the formation of multiple growth facets. This suggests that the attenuation of disorder and/or concentration of structural defects at higher pressures and oxygen flows, can lead to better crystallinity of the anatase phase, regardless of its relative abundance in the $TiO_2$ thin film.

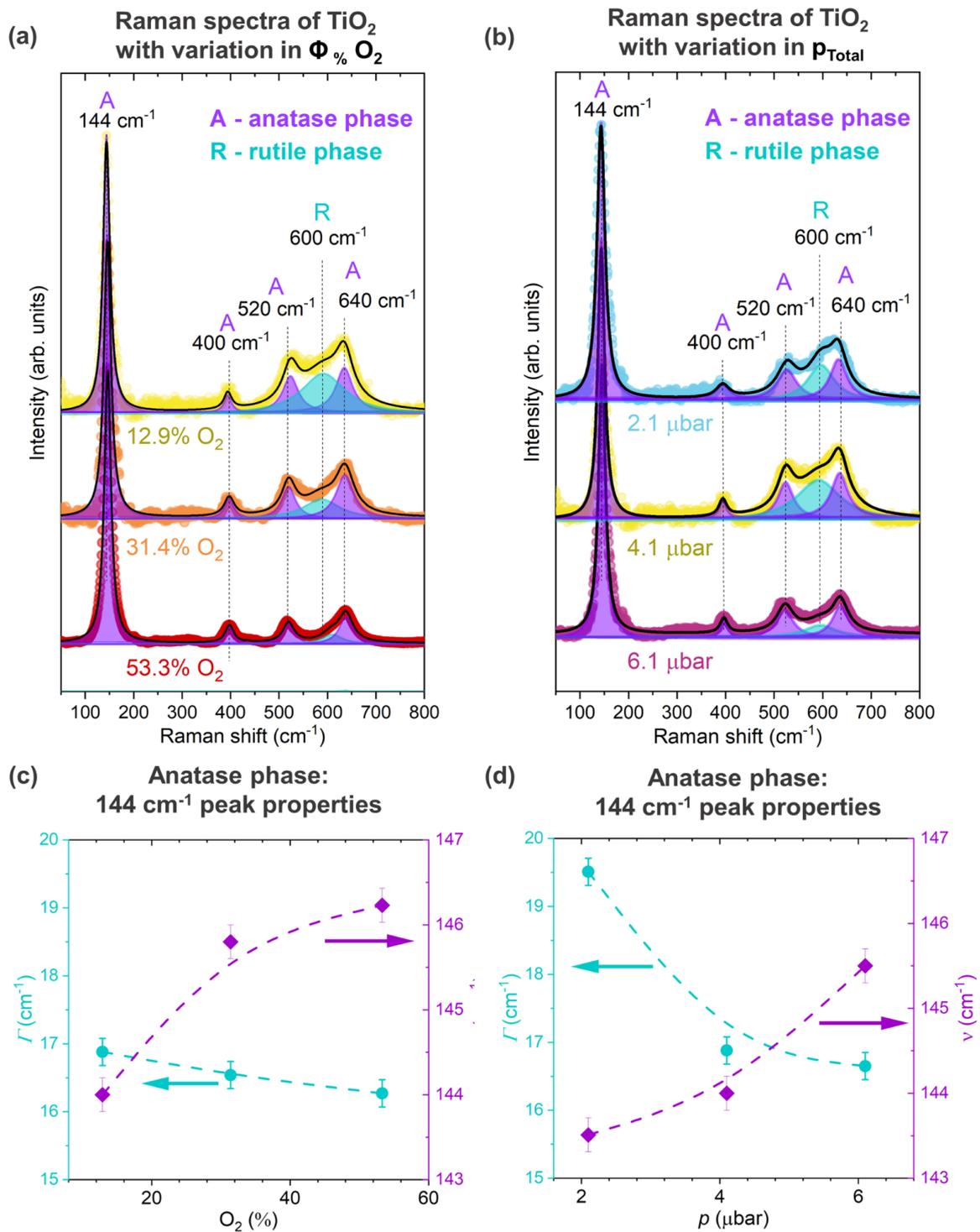

**Figure 7. Raman spectra of TiO$_2$ thin films synthesized by HiPIMS and measured using 532 nm laser excitation**. Raman spectra and peak identification of anatase (labeled A) and rutile (labeled R) phase synthesized for various **(a)** $\Phi_{\%}$ O$_2$ and **(b)** $p_{Total}$. Changes in the width and peak position of the most intense anatase peak located at around 144 cm$^{-1}$ with the variation in **(c)** $\Phi_{\%}$ O$_2$ and **(d)** $p_{Total}$.

Raman spectroscopy measurements were conducted to further verify phase identification and to investigate the crystal quality and presence of defects in the formed phases. **Figure 7a** and **7b** present the Raman spectra of the $TiO_2$ samples synthesized at different $\Phi_\% O_2$ and $p_{Total}$, respectively. Deconvolution of the Raman spectra using Lorentzian curves was performed according to the methodology presented in Ref. [50,51], in order to determine the Raman peak properties and enable phase identification. Most of the Raman peaks observed in the spectra match well with the characteristic peaks of anatase phase positioned at 144 ($E_{1g}$), 400 ($B_{1g}$), 520 ($A_{1g}$), 640 ($E_{3g}$) cm$^{-1}$ [52–55]. Contribution of the rutile phase is observed from the Raman peak at 600 ($A_{1g}$) cm$^{-1}$ [52–56]. Raman peaks from other phases, such as brookite $TiO_2$ were not observed in the Raman spectra, in agreement with the GID results.

To investigate the crystal quality and identify potential defects, we examine the changes in the width and peak position of the most intense Raman mode ($E_{1g}$) of the anatase phase located at around 144 cm$^{-1}$. The width of Raman peak is inversely proportional to the phonon lifetime and sensitive to crystal quality of the material. As the defect concentration in the material increases (i.e. the crystal quality decreases), the phonons become more confined and their lifetime is reduced [57–59]. This leads to an increase in the width of Raman peaks.

The position of Raman modes is also sensitive to crystallinity. For ideal crystals, the Raman mode position is defined as the phonon frequency at the $k = 0$ (Gamma point) in the Brillouin zone, due to the momentum conservation rule. In case of phonon confinement for materials with higher concentration of defects or smaller crystalline domains, the momentum conservation rule is relaxed due to Heisenberg's principle, which can lead to contributions in the Raman spectra from phonons away from the Gamma point [57]. These contributions can move the Raman peak position. The direction of the shift towards lower or higher frequencies will depend on the phonon dispersion of the Raman mode. For the $E_{1g}$ mode of the anatase phase, the phonon dispersion is characterized by lower frequency phonons around the Gamma point [60], which means that the position of $E_{1g}$ mode will shift towards lower frequencies with the increase in defect concentration.

**Figure 7c** and **7d** show the dependence of the width and position of the $E_{1g}$ mode with the change in $\Phi_\% O_2$ and $p_{Total}$, respectively. We observe a reduction in the width and an increase in the peak frequency, signaling a decrease in the defect concentration and improvement in the crystal quality of the anatase phase with both increasing $\Phi_\% O_2$ and deposition pressure.

We can correlate these results with the occurrence of defects in the anatase phase. According to first-principles studies [61], oxygen vacancies ($V_O$) are the most commonly formed intrinsic defects in anatase $TiO_2$ due to their relatively low formation energy compared to other types of point defects. These vacancies are particularly favored under oxygen-deficient synthesis conditions, such as when the oxygen flow ratio ($\Phi_\% O_2$) or the oxygen partial pressure ($pO_2$) is low [62]. In our experiments, the estimated $pO_2$ values varied significantly across the synthesis matrix (details on the calculations are given in the Supporting Information). At the lowest $\Phi_\% O_2$ condition of 8% and total pressure of 0.61 Pa, the oxygen partial pressure is approximately 0.0488 Pa. For 13% $\Phi_\% O_2$ at 0.41 Pa, $pO_2$ is 0.0533 Pa, and for 15% at 0.21 Pa, $pO_2$ is 0.0315 Pa. Along the $\Phi_\% O_2$ axis at fixed pressure (0.41 Pa), increasing the oxygen fraction to 32.4% and 53.3% yields $pO_2$ values of 0.133 Pa and 0.219 Pa, respectively. This nearly sevenfold increase in available oxygen leads to a progressively more oxidizing environment that suppresses the formation of oxygen vacancies. These trends are reflected in the Raman spectra in Figure 7, where higher $\Phi\%\,O_2$ conditions correlate with a narrowing and blue-shifting of the anatase $E_g$ mode—an indicator of reduced defect density and improved crystal order.

The improvement in crystallinity with increasing deposition pressure is also supported by our energetic and structural analysis, despite the relatively narrow variation in $\Phi\%\,O_2$ values along the pressure axis (8–15%, Table 1). As the pressure increases from 0.21 Pa to 0.61 Pa, the mean free path of sputtered species decreases from approximately 32.4 mm to 11.2 mm, increasing the number of gas-phase collisions from ~1.5 to ~4.5. This leads to a marked reduction in the kinetic energy retained by titanium and oxygen species—from ~21% at 0.21 Pa to just ~1% at 0.61 Pa. Under low-pressure conditions, this higher-energy flux facilitates enhanced adatom surface mobility, resulting in more efficient atomic packing and improved crystallization [63]. These energy-driven dynamics also influence the defect landscape: oxygen vacancies are more likely to form under low-pressure, low-$pO_2$ conditions where oxidation is incomplete and mobility is high [64]. In contrast, higher pressures promote a more oxidizing environment—even at constant $\Phi\%\,O_2$—due to the increased absolute oxygen content in the gas phase. This favors the incorporation of oxygen into the $TiO_2$ matrix and supports the formation of dense, well-crystallized anatase with fewer intrinsic defects [65]. Although elevated pressures may enable the stabilization of point defects such as $Ti^{3+}$ or titanium interstitials [66], these are generally less disruptive to the overall crystal structure than oxygen vacancies. These results confirm that increasing deposition pressure, in combination with sufficient oxygen availability, leads to lower defect concentrations and enhanced structural quality in HiPIMS-grown anatase $TiO_2$ thin films. This also demonstrates that HiPIMS can effectively be used to tune the defect concentration in $TiO_2$ films by carefully adjusting both the oxygen flow ratio and total chamber pressure.

**Optoelectronic characterization of $TiO_2$ thin films**

Finally, we investigate the effect of HiPIMS deposition parameters on the optoelectronic properties of $TiO_2$ thin films using photoluminescence (PL) spectroscopy and ellipsometry measurements.

PL properties of $TiO_2$ thin films deposited by HiPIMS under varying oxygen concentrations and total pressures were investigated to elucidate the roles of phase composition, defect chemistry, and electronic recombination pathways. The results are presented in Figure 8. Excitation was performed using a continuous-wave 488 nm (2.54 eV) laser, which is below the fundamental bandgap of both anatase (~3.2 eV) and rutile (~3.0 eV). Under these sub-bandgap excitation conditions, a single, broad PL peak was consistently observed between 2.1 and 2.25 eV (Figure 8). This emission was well-described by a single Gaussian component and exhibited no evidence of higher-energy or near-band-edge transitions. The emission energy range and spectral shape suggest that the radiative process relates to sub-bandgap states, consistent with the so-called "red PL" mechanism previously described in anatase by McHale et al. [67] and elaborated in detail by Pallotti et al. [68] and Mascaretti et al. (2019) [69].

Under sub-bandgap excitation, electrons are promoted from the valence band to deep or intermediate trap states rather than to the conduction band minimum. The observation of PL centered at ~2.1–2.25 eV implies that the involved trap states are located approximately 1.0–1.2 eV below the conduction band edge. These electrons subsequently recombine radiatively with holes residing in the valence band or in shallowly localized levels. The predominance of a single PL band, together with the absence of higher-energy emissions (e.g., from self-trapped excitons or shallow donor-acceptor transitions), indicates that defect-mediated recombination via deep electron traps constitutes the dominant radiative mechanism. Such deep-level traps have been commonly attributed to oxygen vacancies ($V_O$), singly or doubly ionized, and $Ti^{3+}$ centers resulting from localized reduction of $Ti^{4+}$, as substantiated by EPR and XPS analyses in previous studies.

A systematic blueshift of the PL peak, from 2.10 to 2.25 eV, was observed with increasing oxygen concentration at constant pressure. This spectral shift was accompanied by a monotonic decrease in PL intensity. These trends are consistent with a reduced density of deep-level defects under more oxidizing growth conditions. As oxygen availability increases, the formation of $V_O$ and $Ti^{3+}$ centers becomes less favorable, leading to a narrowing and upward shift of the trap energy distribution. The recombination pathway, consequently, shifts toward shallower traps, which emit at higher photon energies. Furthermore, the diminished concentration of radiative centers results in lower PL intensity, as fewer defect sites are available to facilitate radiative transitions. Similar behavior was reported by Mascaretti et al. [69], who showed that thermally reduced anatase $TiO_2$ films exhibited stronger and red-shifted PL (~1.8–2.0 eV) due to abundant deep-level traps, whereas oxidized films displayed broader, weaker, and blueshifted emissions. The emission energy in the present HiPIMS-grown films, positioned between 2.1 and 2.25 eV, suggests a moderately reduced state with a significant yet tunable concentration of electronic defects.

In addition to defect chemistry, phase composition also influences the PL response. GID analysis indicated that anatase is the dominant phase in all films, but its content slightly decreases with increasing oxygen concentration (from 88% to 82%) and increases substantially with pressure (from 75% to 95%) at constant oxygen flow. While both anatase and rutile possess intrinsic defect states, PL in the visible range has been predominantly attributed to anatase. Rutile, by contrast, exhibits weak near-infrared photoluminescence (~1.48–1.51 eV) that becomes apparent under low-temperature or specific excitation conditions, while its emission under room-temperature visible excitation is generally negligible due to efficient non-radiative recombination pathways. Pallotti et al. [68] demonstrated that PL in mixed-phase $TiO_2$ samples is weaker than in pure anatase due to the faster nonradiative recombination kinetics in rutile, which acts as a quenching phase. Specifically, it was shown that the visible PL intensity from anatase-rich films significantly exceeded that of rutile or mixed-phase counterparts under identical excitation conditions. This quenching behavior is attributed to rutile's higher density of fast nonradiative pathways and its reduced capacity to support long-lived excited states that contribute to visible PL.

Despite the modest reduction in anatase content at higher oxygen concentrations, the observed PL suppression is more effectively explained by the reduction in radiative defect density than by phase transformation alone. A 6% change in anatase content is unlikely to produce the observed magnitude of intensity drop unless coupled with chemical changes. However, the increasing rutile content may enhance nonradiative recombination channels, marginally contributing to the overall PL decrease.

On the other hand, when the total pressure was increased at fixed oxygen concentration, anatase content increased markedly (from 75% to 95%), yet the PL peak position remained stable at ~2.1 eV and the intensity exhibited negligible change. This suggest that, although anatase serves as the structural host for PL-active centers, its presence alone is insufficient to determine PL behavior. The chemical nature and occupancy of defect states, particularly deep electron traps such as $Ti^{3+}$ and oxygen vacancies, remain the critical determinants of emission efficiency and spectral character. The minimal effect of pressure on PL implies that the defect chemistry remains relatively unchanged under the tested conditions, consistent with the expectation that oxygen chemical potential — more so than background gas pressure — governs defect formation energetics during film growth.

It is also pertinent to consider the possible influence of a Schottky barrier at the $Ti/TiO_2$ interface on the observed photoluminescence, particularly in light of the sub-bandgap excitation used in this study. Titanium is known to form a rectifying contact with n-type $TiO_2$, leading to the formation of a Schottky barrier and associated depletion region within the semiconductor [70]. The resulting built-in electric field may facilitate carrier separation near the interface, potentially reducing radiative recombination and enhancing nonradiative processes. However, in our film architecture, the $Ti/TiO_2$ interface is located at the bottom of the stack, while optical excitation occurs from the top surface of the $TiO_2$ layer. Under 488 nm excitation (2.54 eV), the optical penetration depth in anatase $TiO_2$ exceeds 1 μm, significantly larger than the total film thickness (~150 nm), meaning the excitation light uniformly penetrates through the entire film thickness. Despite the potential for a depletion region extending tens of nanometers from the buried Ti interface, the majority of photogenerated carriers — and thus the dominant PL signal — originate in the electrically neutral bulk and near-surface regions of the film. Given this excitation geometry and the consistent PL behavior observed across samples with varying structural properties, the contribution of the buried Schottky interface to the steady-state PL signal is expected to be negligible. These considerations suggest that the emission characteristics are primarily governed by bulk or near-surface recombination processes associated with intrinsic defect states, rather than being significantly influenced by interfacial effects.

In summary, the PL response of HiPIMS-grown $TiO_2$ thin films under sub-bandgap excitation is dominated by radiative recombination at deep defect states in the anatase phase, specifically involving $Ti^{3+}$ centers and oxygen vacancies. Increasing the oxygen content during synthesis was found to suppress these defects, leading to a blueshift and reduction in PL intensity. In contrast, variations in total pressure predominantly affected phase composition, enhancing anatase fraction without significantly impacting the PL features, suggesting that structural phase alone does not control optical activity. These results reinforce the interpretation that PL in $TiO_2$ is primarily governed by defect chemistry rather than crystallographic phase, and that sub-bandgap steady-state PL spectroscopy serves as an effective probe of defect states in mixed-phase oxide thin films.

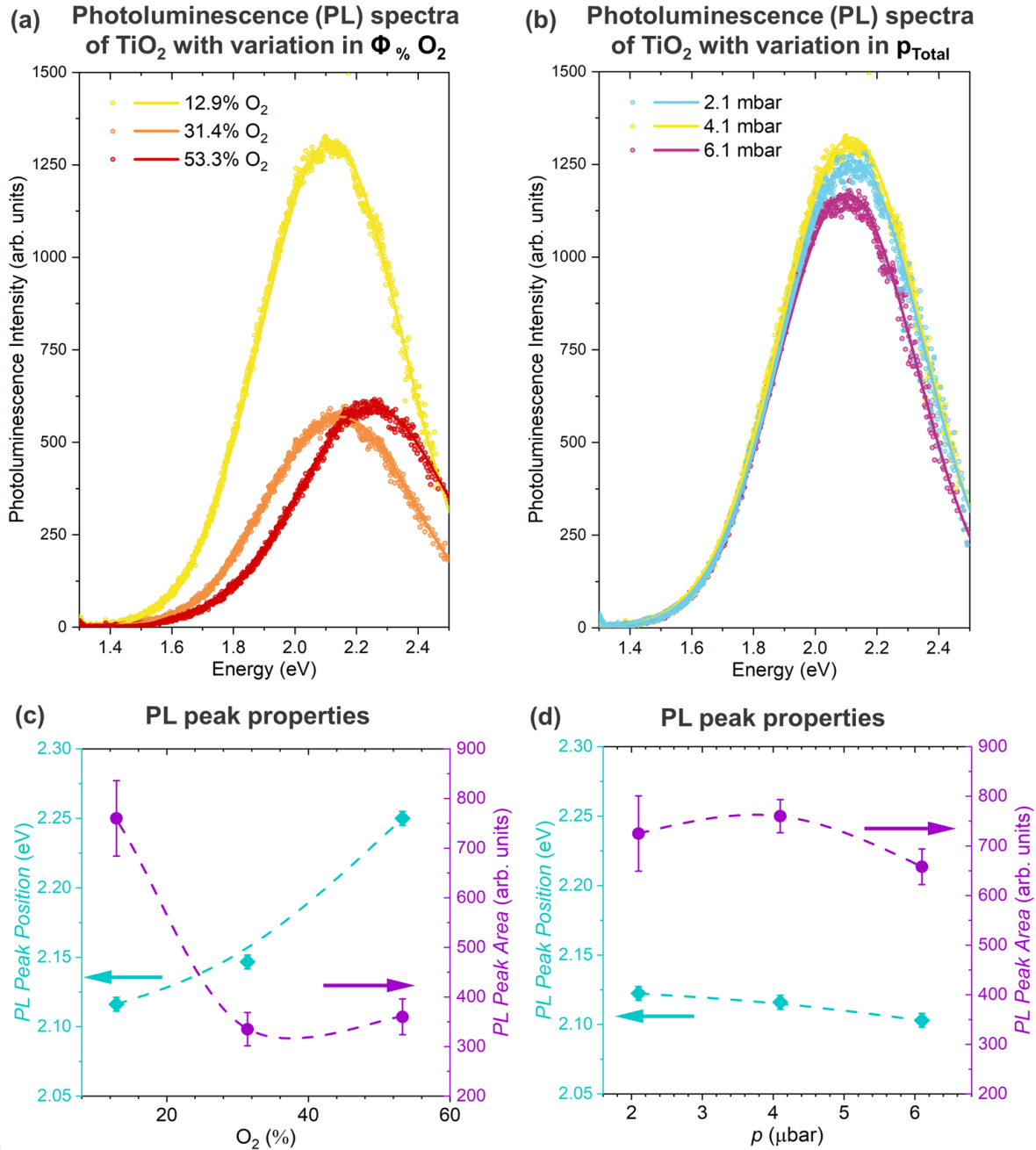

**Figure 8. Photoluminescence spectra of TiO$_2$ films synthesized by HiPIMS and measured with 488 nm laser excitation**. PL spectra of TiO$_2$ films synthesized for various **(a)** $\Phi_{\%}$ O$_2$ and **(b)** $p_{\text{Total}}$. PL peak properties (peak position and peak area) plotted in dependence of **(c)** $\Phi_{\%}$ O$_2$ and **(d)** deposition pressure.

Variations in the refractive index of the TiO$_2$ films were investigated using ellipsometry measurements. **Figure 9a** and **9b** present changes in the refractive index of TiO$_2$ films synthesized for various $\Phi_{\%}$ O$_2$ and $p_{\text{Total}}$, respectively. Overall, the measured refractive indices are in good agreement with those reported in the literature for mixed-phase anatase and rutile TiO$_2$ films [71–73]. We observe a general rise in refractive index with increasing $\Phi_{\%}$ O$_2$. This can be attributed to the changes in the ratio of anatase and rutile phase present in the TiO$_2$ film. From GID measurements, we observed a higher quantity of rutile phase for increased $\Phi_{\%}$ $O_2$. As the rutile phase has a higher refractive index when compared to anatase[71–73], the refractive index of polycrystalline $TiO_2$ films can correspondingly be expected to rise when the rutile fraction therein increases. Likewise, the lowest deposition pressure, with the strongest rutile contribution in our GID results, has the highest refractive index, reflecting the higher proportion of rutile in HiPIMS $TiO_2$ synthesized at this pressure, relative to the films synthesized at the higher pressures. Not many differences are observed between the intermediate and highest-pressure conditions. This is probably due to the interplay between the phase composition and the crystallinity of the phases, which can both affect the overall refractive index. These results point to the possibility of tuning of the refractive index through variation in oxygen flow during synthesis.

It should be noted that these refractive index changes should not be related to poor stoichiometry or low film quality. RBS and HI-ERDA analyses (Table 2) confirm that all films are nearly stoichiometric, with O:Ti atomic ratios close to 2.00 across all conditions. The observed optical trends are instead a direct consequence of controlled variations in phase composition and film microstructure. This is further supported by Raman spectroscopy, which shows a narrowing and blue-shifting of the dominant anatase Eg peak at higher $\Phi$ % $O_2$ and lower pressure—indicative of reduced defect concentration and improved crystallinity (Figures 6 and 7). Therefore, the refractive index evolution in these films reflects a well-regulated structural tuning rather than any degradation or stoichiometric inconsistency.

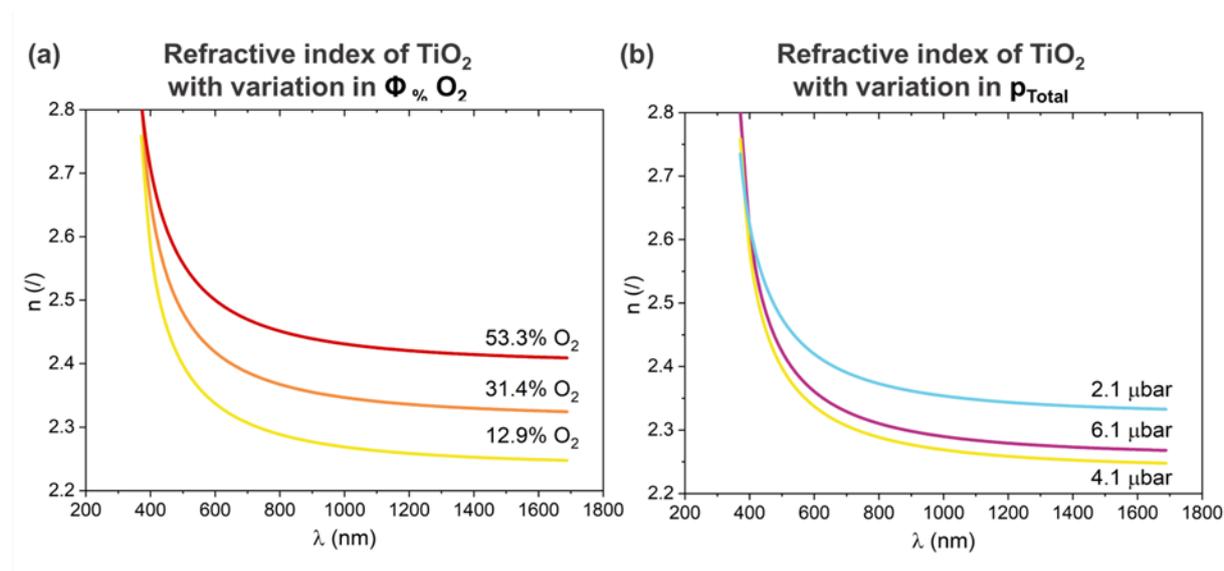

**Figure 9. Optical characterization of $TiO_2$ films using ellipsometry measurements**. Refractive index of $TiO_2$ films synthesized by HiPIMS at various **(a)** $\Phi_{\%}$ $O_2$ and **(b)** $p_{Total}$.

**Conclusions**

Our work explored the properties of low temperature HiPIMS $TiO_2$ thin films deposited over a wide range of $\Phi_{\%}$ $O_2$ and $p_{Total}$ growth parameters, through which we could tune the titania polymorph composition, and with it, the relevant structural and optoelectronic properties. All films showed a nearly stoichiometric composition and very similar morphological properties, while a significant dependence of the rutile and anatase phase composition on the deposition parameters was observed by GID and Raman spectroscopy. The formation of the rutile phase is favored by high $\Phi_\%$ $O_2$ and low $p_{Total}$, while a low $\Phi_\%$ $O_2$ and high $p_{Total}$ increase the anatase content. LeBail refinements revealed the rutile phase to be robust against changes in its crystal structure, while a variation of the lattice constant was observed for the anatase and attributed to a drop in oxygen vacancy concentration with increased $\Phi_\%$ $O_2$. In terms of optical properties, increasing the $\Phi_\%$ $O_2$ resulted in increased refractive indices and a decrease in photoluminescence originated from the bands. Both properties seemed less sensitive with variations in $p_{Total}$. We explain these observations with quantity changes in rutile and anatase phases present in the differently grown $TiO_2$ films, as well as variation in crystallinity and oxygen vacancy concentration. Our results demonstrate that the HiPIMS method enables tuning of $TiO_2$ thin films properties that extend beyond polymorph-composition control and into defect engineering, useful in diverse applications, such as photovoltaics, sensors, and photocatalysis. This adds to HiPIMS's critically important ability to deposit high-quality $TiO_2$ films on temperature-sensitive substrates, and in so doing, enhances the deployment potential of this versatile material in the advanced technologies and applications of tomorrow.

**Author contributions**

A.C., O.Y., H.J.H and M.D. conceived the research idea. H.J.H and M.D. supervised the work. A.C. prepared the samples, and performed AFM measurements. E.H. performed ellipsometry measurements and analyzed the data. S.L. prepared lamellas and acquired S/TEM images. R.C. performed SEM imaging. A.M. did compositional analysis using RBS and HI-ERDA. M.D. did Raman and PL measurements. A.C. and M.D. analyzed all the data. A.C. and M.D. wrote the manuscript with input from all co-authors.

**Competing Interests**

The authors declare that they have no competing financial interests.

**References**

[1]  Li Z, Li Z, Zuo C and Fang X 2022 Application of Nanostructured TiO2 in UV Photodetectors: A Review *Advanced Materials* **34** 2109083

[2]  Bang S Y, Fan X-B, Jung S-M, Yang J, Shin D-W, Suh Y-H, Lee T H, Lee S, Choi H W, Occhipinti L G, Han S D and Kim J M 2020 Highly Stable and Scalable Blue QD-LED via an Evaporated TiO2 Thin Film as an Electron Transport Layer *Advanced Optical Materials* **8** 2001172

[3]  Crossland E J W, Noel N, Sivaram V, Leijtens T, Alexander-Webber J A and Snaith H J 2013 Mesoporous TiO2 single crystals delivering enhanced mobility and optoelectronic device performance *Nature* **495** 215–9

[4] Baldini E, Chiodo L, Dominguez A, Palummo M, Moser S, Yazdi-Rizi M, Auböck G, Mallett B P P, Berger H, Magrez A, Bernhard C, Grioni M, Rubio A and Chergui M 2017 Strongly bound excitons in anatase TiO2 single crystals and nanoparticles *Nat Commun* **8** 13

[5] Ding Y, Ding B, Kanda H, Usiobo O J, Gallet T, Yang Z, Liu Y, Huang H, Sheng J, Liu C, Yang Y, Queloz V I E, Zhang X, Audinot J-N, Redinger A, Dang W, Mosconic E, Luo W, De Angelis F, Wang M, Dörflinger P, Armer M, Schmid V, Wang R, Brooks K G, Wu J, Dyakonov V, Yang G, Dai S, Dyson P J and Nazeeruddin M K 2022 Single-crystalline TiO2 nanoparticles for stable and efficient perovskite modules *Nat. Nanotechnol.* **17** 598–605

[6] Kumar N, Patel M, Nguyen T T, Kim S and Kim J 2021 Effect of TiO2 layer thickness of TiO2/NiO transparent photovoltaics *Progress in Photovoltaics: Research and Applications* **29** 943–52

[7] Ge M, Cai J, Iocozzia J, Cao C, Huang J, Zhang X, Shen J, Wang S, Zhang S, Zhang K-Q, Lai Y and Lin Z 2017 A review of TiO2 nanostructured catalysts for sustainable H2 generation *International Journal of Hydrogen Energy* **42** 8418–49

[8] Guo Q, Zhou C, Ma Z and Yang X 2019 Fundamentals of TiO2 Photocatalysis: Concepts, Mechanisms, and Challenges *Advanced Materials* **31** 1901997

[9] Tomás-Gamasa M and Mascareñas J L 2020 TiO2-Based Photocatalysis at the Interface with Biology and Biomedicine *ChemBioChem* **21** 294–309

[10] Wu A and Ren W 2020 TiO2 Nanoparticles: Applications in Nanobiotechnology and Nanomedicine *John Wiley & Sons*

[11] Jafari S, Mahyad B, Hashemzadeh H, Janfaza S, Gholikhani T and Tayebi L 2020 Biomedical Applications of TiO2 Nanostructures: Recent Advances *International Journal of Nanomedicine* **15** 3447–70

[12] Zhou C, Xi Z, Stacchiola D J and Liu M 2022 Application of ultrathin TiO2 layers in solar energy conversion devices *Energy Science & Engineering* **10** 1614–29

[13] Fatima Q, Haidry A A, Zhang H, El Jery A and Aldrdery M 2024 A critical review on advancement and challenges in using TiO2 as electron transport layer for perovskite solar cell *Materials Today Sustainability* **27** 100857

[14] Zhang P, Lu X F, Luan D and Lou X W (David) 2020 Fabrication of Heterostructured Fe2TiO5–TiO2 Nanocages with Enhanced Photoelectrochemical Performance for Solar Energy Conversion *Angewandte Chemie* **132** 8205–9

[15] Zhang W, He H, Li H, Duan L, Zu L, Zhai Y, Li W, Wang L, Fu H and Zhao D 2021 Visible-Light Responsive TiO2-Based Materials for Efficient Solar Energy Utilization *Advanced Energy Materials* **11** 2003303

[16] Zhang H and Banfield J F 2014 Structural Characteristics and Mechanical and Thermodynamic Properties of Nanocrystalline TiO2 *Chem. Rev.* **114** 9613–44

[17] Hanaor D A H and Sorrell C C 2011 Review of the anatase to rutile phase transformation *J Mater Sci* **46** 855–74

[18] Xu X, Fukuda K, Karki A, Park S, Kimura H, Jinno H, Watanabe N, Yamamoto S, Shimomura S, Kitazawa D, Yokota T, Umezu S, Nguyen T-Q and Someya T 2018 Thermally stable, highly

efficient, ultraflexible organic photovoltaics *Proceedings of the National Academy of Sciences* **115** 4589–94

[19] Liu J, Zheng D, Wang K, Li Z, Liu S, Peng L and Yang D 2024 Evolutionary manufacturing approaches for advancing flexible perovskite solar cells *Joule* **8** 944–69

[20] Qadir A, Shafique S, Iqbal T, Ali H, Xin L, Ruibing S, Shi T, Xu H, Wang Y and Hu Z 2024 Recent advancements in polymer-based photodetector: A comprehensive review *Sensors and Actuators A: Physical* **370** 115267

[21] Zhang X, Zhang S, Cui X, Zhou W, Cao W, Cheng D and Sun Y 2022 Recent Advances in TiO2-based Photoanodes for Photoelectrochemical Water Splitting *Chemistry – An Asian Journal* **17** e202200668

[22] Chen J, Zhang J, Ye M, Rao Z, Tian T, Shu L, Lin P, Zeng X and Ke S 2020 Flexible TiO2/Au thin films with greatly enhanced photocurrents for photoelectrochemical water splitting *Journal of Alloys and Compounds* **815** 152471

[23] Subudhi P and Punetha D 2023 Progress, challenges, and perspectives on polymer substrates for emerging flexible solar cells: A holistic panoramic review *Progress in Photovoltaics: Research and Applications* **31** 753–89

[24] West G, Kelly P, Barker P, Mishra A and Bradley J 2009 Measurements of Deposition Rate and Substrate Heating in a HiPIMS Discharge *Plasma Processes and Polymers* **6** S543–7

[25] Kelly P J, Barker P M, Ostovarpour S, Ratova M, West G T, Iordanova I and Bradley J W 2012 Deposition of photocatalytic titania coatings on polymeric substrates by HiPIMS *Vacuum* **86** 1880–2

[26] Straňák V, Quaas M, Wulff H, Hubička Z, Wrehde S, Tichý M and Hippler R 2008 Formation of TiOx films produced by high-power pulsed magnetron sputtering *J. Phys. D: Appl. Phys.* **41** 055202

[27] Straňák V, Čada M, Quaas M, Block S, Bogdanowicz R, Kment Š, Wulff H, Hubička Z, Helm C A, Tichý M and Hippler R 2009 Physical properties of homogeneous TiO2 films prepared by high power impulse magnetron sputtering as a function of crystallographic phase and nanostructure *J. Phys. D: Appl. Phys.* **42** 105204

[28] Alami J, Sarakinos K, Uslu F, Klever C, Dukwen J and Wuttig M 2009 On the phase formation of titanium oxide films grown by reactive high power pulsed magnetron sputtering *J. Phys. D: Appl. Phys.* **42** 115204

[29] Amin A, Köhl D and Wuttig M 2010 The role of energetic ion bombardment during growth of TiO2 thin films by reactive sputtering *J. Phys. D: Appl. Phys.* **43** 405303

[30] Aiempanakit M, Helmersson U, Aijaz A, Larsson P, Magnusson R, Jensen J and Kubart T 2011 Effect of peak power in reactive high power impulse magnetron sputtering of titanium dioxide *Surface and Coatings Technology* **205** 4828–31

[31] Konstantinidis S, Dauchot J P and Hecq M 2006 Titanium oxide thin films deposited by high-power impulse magnetron sputtering *Thin Solid Films* **515** 1182–6

[32] Cemin F, Tsukamoto M, Keraudy J, Antunes V G, Helmersson U, Alvarez F, Minea T and Lundin D 2018 Low-energy ion irradiation in HiPIMS to enable anatase TiO2 selective growth *J. Phys. D: Appl. Phys.* **51** 235301

[33]   Nouvellon C, Michiels M, Dauchot J P, Archambeau C, Laffineur F, Silberberg E, Delvaux S, Cloots R, Konstantinidis S and Snyders R 2012 Deposition of titanium oxide films by reactive High Power Impulse Magnetron Sputtering (HiPIMS): Influence of the peak current value on the transition from metallic to poisoned regimes *Surface and Coatings Technology* **206** 3542–9

[34]   Ratova M, West G T and Kelly P J 2014 Optimisation of HiPIMS photocatalytic titania coatings for low temperature deposition *Surface and Coatings Technology* **250** 7–13

[35]   Surpi A, Kubart T, Giordani D, Tosello M, Mattei G, Colasuonno M and Patelli A 2013 HiPIMS deposition of TiOx in an industrial-scale apparatus: Effects of target size and deposition geometry on hysteresis *Surface and Coatings Technology* **235** 714–9

[36]   Barker P M, Lewin E and Patscheider J 2013 Modified high power impulse magnetron sputtering process for increased deposition rate of titanium *Journal of Vacuum Science & Technology A* **31** 060604

[37]   Doolittle L R 1985 Algorithms for the rapid simulation of Rutherford backscattering spectra *Nuclear Instruments and Methods in Physics Research Section B: Beam Interactions with Materials and Atoms* **9** 344–51

[38]   Döbeli M, Kottler C, Stocker M, Weinmann S, Synal H-A, Grajcar M and Suter M 2004 Gas ionization chambers with silicon nitride windows for the detection and identification of low energy ions *Nuclear Instruments and Methods in Physics Research Section B: Beam Interactions with Materials and Atoms* **219–220** 415–9

[39]   Arstila K, Julin J, Laitinen M I, Aalto J, Konu T, Kärkkäinen S, Rahkonen S, Raunio M, Itkonen J, Santanen J-P, Tuovinen T and Sajavaara T 2014 Potku – New analysis software for heavy ion elastic recoil detection analysis *Nuclear Instruments and Methods in Physics Research Section B: Beam Interactions with Materials and Atoms* **331** 34–41

[40]   Dimitrievska M, Fairbrother A, Gunder R, Gurieva G, Xie H, Saucedo E, Pérez-Rodríguez A, Izquierdo-Roca V and Schorr S 2016 Role of S and Se atoms on the microstructural properties of kesterite $Cu_2ZnSn(S_xSe_{1-x})_4$ thin film solar cells *Physical Chemistry Chemical Physics* **18** 8692–700

[41]   Gates-Rector S and Blanton T 2019 The Powder Diffraction File: a quality materials characterization database *Powder Diffraction* **34** 352–60

[42]   Mardare D, Tasca M, Delibas M and Rusu G I 2000 On the structural properties and optical transmittance of TiO2 r.f. sputtered thin films *Applied Surface Science* **156** 200–6

[43]   Spurr R A and Myers Howard 1957 Quantitative Analysis of Anatase-Rutile Mixtures with an X-Ray Diffractometer *Anal. Chem.* **29** 760–2

[44]   Ramakrishnan R, Kalaivani S, Amala Infant Joice J and Sivakumar T 2012 Photocatalytic activity of multielement doped TiO2 in the degradation of congo red *Applied Surface Science* **258** 2515–21

[45]   Manuputty M Y, Lindberg C S, Dreyer J A H, Akroyd J, Edwards J and Kraft M 2021 Understanding the anatase-rutile stability in flame-made $TiO_2$ *Combustion and Flame* **226** 347–61

[46]   Samin A J 2021 Oxidation thermodynamics of Nb-Ti alloys studied via first-principles calculations *Journal of Alloys and Compounds* **879** 160455

[47]   Agnarsson B, Magnus F, Tryggvason T K, Ingason A S, Leosson K, Olafsson S and Gudmundsson J T 2013 Rutile TiO2 thin films grown by reactive high power impulse magnetron sputtering *Thin Solid Films* **545** 445–50

[48]   Yang Y-J, Tsou H-K, Chen Y-H, Chung C-J and He J-L 2015 Enhancement of bioactivity on medical polymer surface using high power impulse magnetron sputtered titanium dioxide film *Materials Science and Engineering: C* **57** 58–66

[49]   He T, Wang D, Xu Y and Zhang J 2024 The Facile Construction of Anatase Titanium Dioxide Single Crystal Sheet-Connected Film with Observable Strong White Photoluminescence *Coatings* **14** 292

[50]   Blaga C, Álvarez Á L, Balgarkashi A, Banerjee M, Morral A F i and Dimitrievska M 2024 Unveiling the complex phonon nature and phonon cascades in 1L to 5L WSe 2 using multiwavelength excitation Raman scattering *Nanoscale Advances* **6** 4591–603

[51]   Dimitrievska M, Litvinchuk A P, Zakutayev A and Crovetto A 2023 Phonons in Copper Diphosphide (CuP2): Raman Spectroscopy and Lattice Dynamics Calculations *J. Phys. Chem. C* **127** 10649–54

[52]   Meinhold G 2010 Rutile and its applications in earth sciences *Earth-Science Reviews* **102** 1–28

[53]   Lubas M, Jasinski J J, Sitarz M, Kurpaska L, Podsiad P and Jasinski J 2014 Raman spectroscopy of TiO2 thin films formed by hybrid treatment for biomedical applications *Spectrochimica Acta Part A: Molecular and Biomolecular Spectroscopy* **133** 867–71

[54]   Zhang J, Li M, Feng Z, Chen J and Li C 2006 UV Raman Spectroscopic Study on TiO2. I. Phase Transformation at the Surface and in the Bulk *J. Phys. Chem. B* **110** 927–35

[55]   Li W, Liang R, Hu A, Huang Z and Zhou Y N 2014 Generation of oxygen vacancies in visible light activated one-dimensional iodine TiO2 photocatalysts *RSC Adv.* **4** 36959–66

[56]   Zhao X, Jin W, Cai J, Ye J, Li Z, Ma Y, Xie J and Qi L 2011 Shape- and Size-Controlled Synthesis of Uniform Anatase TiO2 Nanocuboids Enclosed by Active 100 and 001 Facets *Advanced Functional Materials* **21** 3554–63

[57]   Dimitrievska M, Fairbrother A, Pérez-Rodríguez A, Saucedo E and Izquierdo-Roca V 2014 Raman scattering crystalline assessment of polycrystalline Cu2ZnSnS4 thin films for sustainable photovoltaic technologies: Phonon confinement model *Acta Materialia* **70** 272–80

[58]   Dimitrievska M, Fairbrother A, Saucedo E, Pérez-Rodríguez A and Izquierdo-Roca V 2015 Influence of compositionally induced defects on the vibrational properties of device grade Cu2ZnSnSe4 absorbers for kesterite based solar cells *Applied Physics Letters* **106** 073903

[59]   Dimitrievska M, Oliva F, Guc M, Giraldo S, Saucedo E, Pérez-Rodríguez A and Izquierdo-Roca V 2019 Defect characterisation in Cu 2 ZnSnSe 4 kesterites via resonance Raman spectroscopy and the impact on optoelectronic solar cell properties *Journal of Materials Chemistry A* **7** 13293–304

[60]   Ghose K K, Liu Y and Frankcombe T J 2023 Comparative first-principles structural and vibrational properties of rutile and anatase TiO2 *J. Phys.: Condens. Matter* **35** 505702

[61] Na-Phattalung S, Smith M F, Kim K, Du M-H, Wei S-H, Zhang S B and Limpijumnong S 2006 First-principles study of native defects in anatase Ti O 2 *Phys. Rev. B* **73** 125205

[62] Song W, Jiang Q, Xie X, Brookfield A, McInnes E J L, Shearing P R, Brett D J L, Xie F and Riley D J 2019 Synergistic storage of lithium ions in defective anatase/rutile TiO2 for high-rate batteries *Energy Storage Materials* **22** 441–9

[63] Liu Y, Huang H M and Lin X D 2013 Structural and Optical Properties of Pulse Laser Deposited TiO2 Thin Films *Key Engineering Materials* **537** 224–8

[64] Pan X, Yang M-Q, Fu X, Zhang N and Xu Y-J 2013 Defective TiO 2 with oxygen vacancies: synthesis, properties and photocatalytic applications *Nanoscale* **5** 3601–14

[65] Kang X, Dong G and Dong T 2023 Oxygen Vacancy Defect Engineering of Heterophase Junction TiO2: Interfacial/Surface Oxygen Vacancies Coadjust the Photocatalytic ROS Production *ACS Appl. Energy Mater.* **6** 1025–36

[66] Patra S, Davoisne C, Bruyère S, Bouyanfif H, Cassaignon S, Taberna P-L and Sauvage F 2013 Room-Temperature Synthesis of High Surface Area Anatase TiO2 Exhibiting a Complete Lithium Insertion Solid Solution *Particle & Particle Systems Characterization* **30** 1093–104

[67] Rich C C, Knorr F J and McHale J L 2010 Trap State Photoluminescence of Nanocrystalline and Bulk TiO2,: Implications for Carrier Transport *MRS Online Proceedings Library* **1268** 308

[68] Pallotti D K, Passoni L, Maddalena P, Di Fonzo F and Lettieri S 2017 Photoluminescence Mechanisms in Anatase and Rutile TiO2 *J. Phys. Chem. C* **121** 9011–21

[69] Mascaretti L, Russo V, Zoppellaro G, Lucotti A, Casari C S, Kment Š, Naldoni A and Li Bassi A 2019 Excitation Wavelength- and Medium-Dependent Photoluminescence of Reduced Nanostructured TiO2 Films *J. Phys. Chem. C* **123** 11292–303

[70] Yu H, Schaekers M, Schram T, Demuynck S, Horiguchi N, Barla K, Collaert N, Thean A V-Y and De Meyer K 2016 Thermal Stability Concern of Metal-Insulator-Semiconductor Contact: A Case Study of Ti/TiO2/n-Si Contact *IEEE Transactions on Electron Devices* **63** 2671–6

[71] Bendavid A, Martin P J and Takikawa H 2000 Deposition and modification of titanium dioxide thin films by filtered arc deposition *Thin Solid Films* **360** 241–9

[72] Wojcieszak D, Mazur M, Kaczmarek D, Poniedziałek A, Domanowski P, Szponar B, Czajkowska A and Gamian A 2016 Effect of the structure on biological and photocatalytic activity of transparent titania thin-film coatings *Materials Science-Poland* **34** 856–62

[73] Kang M, Kim S W and Park H Y 2018 Optical properties of TiO2 thin films with crystal structure *Journal of Physics and Chemistry of Solids* **123** 266–70

# Supporting Information

*for:*

**On Tailoring Structural and Optoelectronic Properties of TiO$_2$ Thin Films Synthesized via "Room" Temperature High Power Impulse Magnetron Sputtering (HiPIMS)**

Aarati Chacko,[1] Erwin Hack,[2] Sebastian Lohde,[2] Robin Bucher,[2] Oguz Yildirim,[1] Arnold Mueller,[3] Michel Calame,[2,4,5] Hans J. Hug,[1,5] Mirjana Dimitrievska[2*]

*1 – Functional and Magnetic Thin Films, Swiss Federal Laboratories for Material Science and Technology (EMPA) Ueberlandstrasse 129, 8600 Duebendorf, Switzerland*

*2 - Transport at Nanoscale Interfaces Laboratory, Swiss Federal Laboratories for Material Science and Technology (EMPA) Ueberlandstrasse 129, 8600 Duebendorf, Switzerland*

*3 - Laboratory of Ion Beam Physics, ETH Zurich, CH-8093 Zurich, Switzerland*

*4 – Swiss Nanoscience Institute, University of Basel, 4056 Basel, Switzerland*

*5 – Department of Physics, University of Basel, CH-4056, Basel, Switzerland*

*\*corresponding author:* mirjana.dimitrievska@empa.ch

*SEM and AFM topographies, increasing pressure*

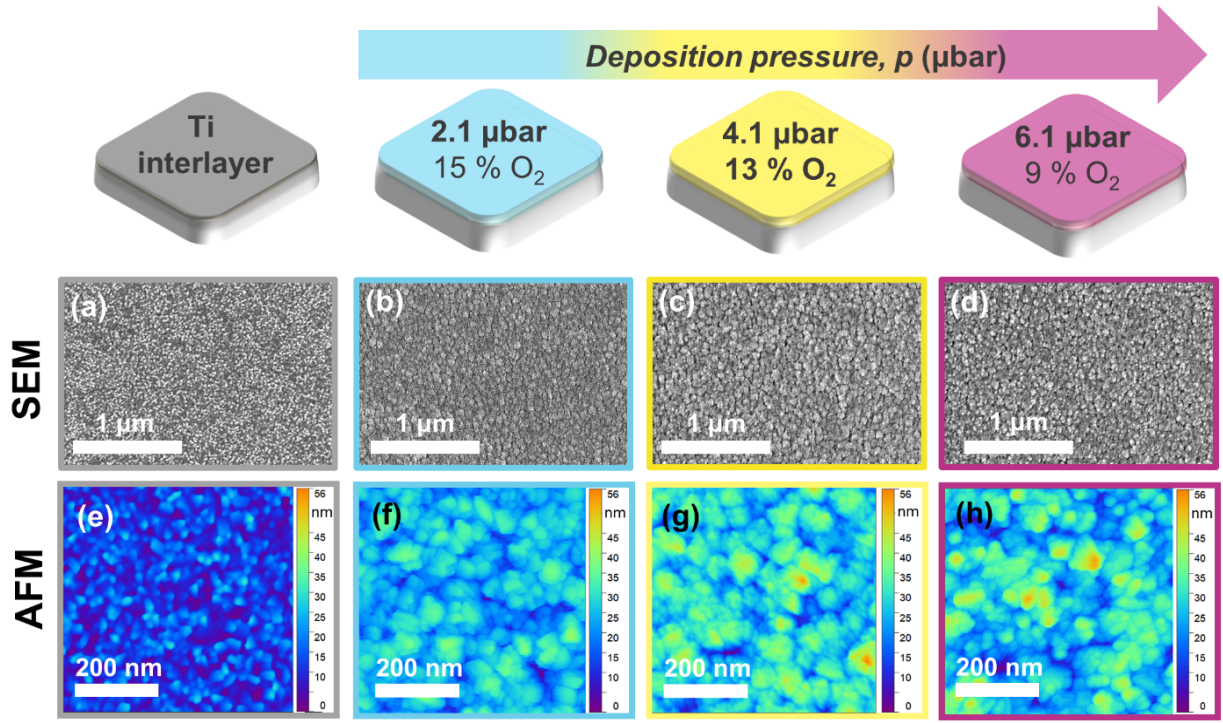

**Figure S1.** Morphological assessment of Ti interlayer and TiO$_2$ thin films deposited for various deposition pressures. **(a) – (d)** SEM and **(e) – (h)** AFM top view images of the Ti interlayer and TiO$_2$ thin films.

*RBS and Hi-ERDA sample composition*

RBS and HI-ERDA are complementary measurements that were used to evaluate the $TiO_2$ film compositions. RBS is non-destructive, that has a depth resolution of the order of the $TiO_2$ films, and is fairly precise, with an error in the atomic% below 5% for our coatings. Nevertheless, RBS cannot detect hydrogen. In contrast, HI-ERDA, is highly sensitive to the light elements, including H, C and O. It is, however, less precise with the estimated error between 15 and 20% for the values in the above table. HI-ERDA is moreover, established through a sputter profile, so some amount of atomic mixing can be expected. This is why we are not surprised to see differences in atomic composition values between RBS and HI-ERDA, and why we lend more credence to the accuracy and precision of the RBS. The HI-ERDA, critically, enabled us to identifying the C and H in our $TiO_2$ films, and we get an order of magnitude for the relative abundance of these contaminants.

**Table S1. HI-ERDA atomic composition of the $TiO_2$ thin films grown under different pressure and %flows of $O_2$ conditions**

| $TiO_2$ synthesis conditions | RBS | | HI-ERDA | | | |
| --- | --- | --- | --- | --- | --- | --- |
| | *Film* | | *Film* | | *Contaminants* | |
| | Ti % | O % | Ti % | O % | C % | H % |
| 2.1 µbar, 15 %$O_2$ | 33±4 | 67±4 | 29±5 | 62±23 | 1.20±0.01 | 7.9±0.4 |
| 4.1 µbar, 13 %$O_2$ | 33±3 | 67±3 | 28±5 | 59±21 | 4.80±0.14 | 7.7±0.4 |
| 4.1 µbar, 31 %$O_2$ | 33±3 | 67±3 | 29±5 | 59±21 | 3.70±0.09 | 8.9±0.5 |
| 4.1 µbar, 53 %$O_2$ | 34±4 | 66±3 | 30±5 | 63±23 | 1.50±0.01 | 5.5±0.2 |
| 6.1 µbar, 8 %$O_2$ | 33±3 | 67±3 | 27±5 | 54±19 | 5.90±0.23 | 13.2±1.1 |

*Estimation of Mean Free Path and Energy Loss during Deposition*

To quantify the impact of pressure on the kinetic energy of sputtered species during film growth, we estimated the mean free path ($\lambda$) and relative energy retention ($E/E_0$) of sputtered particles using standard kinetic gas theory.

*Mean Free Path Calculation*

The mean free path of a sputtered species (in this case, Ti in Ar) is given by:

$$\lambda = \frac{kT}{\sqrt{2}\,\pi\,d^2\,p}$$

Where $k = 1.38 \times 10^{-23}$ J/K (Boltzmann constant), $T = 300$ K (room temperature), $d = 3.7 \times 10^{-10}$ m (effective diameter of Ar atoms) and $p$ is the deposition pressure in Pascals.

Using this relation, we calculated $\lambda$ for the three pressures used in our study: 0.21 Pa, 0.41 Pa, and 0.61 Pa.

*Energy Attenuation Model*

We assume that the sputtered atoms lose energy through gas-phase collisions and that this energy loss follows an exponential decay as a function of the number of collisions:

$$E = E_0\, e^{-\frac{L}{\lambda}}$$

Where $E$ is the retained energy upon substrate arrival, $E_0$ is the initial kinetic energy of the sputtered particle, $L = 0.05$ m is the estimated target-to-substrate distance and $\lambda$ is the calculated mean free path.

Table S2. Calculated Mean Free Path and Energy Loss during Deposition

| Pressure $p$ (Pa) | Mean Free Path $\lambda$ (mm) | Collisions ($L/\lambda$) | Relative Energy ($E/E_0$) |
|---|---|---|---|
| **0.21** | 32.4 | 1.54 | 0.21 (~21%) |
| **0.41** | 16.6 | 3.01 | 0.049 (~5%) |
| **0.61** | 11.2 | 4.48 | 0.011 (~1.1%) |

These calculations clearly demonstrate that increasing pressure significantly reduces the mean free path and increases the number of collisions, resulting in a steep decline in the kinetic energy of arriving species. This supports our interpretation that lower-pressure conditions enable higher-energy flux to the substrate, favoring dense rutile phase formation, while higher pressures lead to energy-dissipative growth conditions more conducive to anatase formation and greater defect accommodation.

*Estimation of Oxygen Partial Pressure (pO₂)*

To assess the chemical environment during TiO$_2$ film growth, the oxygen partial pressure (pO$_2$) was estimated from the total deposition pressure (pTotal) and the oxygen flow fraction (Φ % O$_2$) used in each synthesis condition. Assuming ideal gas mixing, the oxygen partial pressure was calculated using the relation:

$$p_{O_2} = \Phi\%O_2 \cdot p_{total}$$

Where pO$_2$ is the partial pressure of oxygen in Pascals (Pa), Φ % O$_2$ is the oxygen flow ratio (expressed as a decimal), pTotal is the total chamber pressure during deposition (in Pa).

This calculation provides a quantitative estimate of the oxygen availability under each condition and is useful for understanding its impact on film stoichiometry, phase formation, and defect concentration.

**Table S3. Calculated Oxygen Partial Pressures**

| Φ % O₂ | pTotal (Pa) | Calculated pO₂ (Pa) |
|---|---|---|
| 15% | 0.21 | 0.0315 |
| 13% | 0.41 | 0.0533 |
| 8% | 0.61 | 0.0488 |
| 32.4% | 0.41 | 0.1328 |
| 53.3% | 0.41 | 0.2185 |

*Ellipsometry analysis*

Spectroscopic ellipsometry data were analyzed using a multi-layer optical model constructed in CompleteEASE (J.A. Woollam Co.). The $TiO_2$ films were modeled using a Kramers-Kronig-consistent Tauc-Lorentz dispersion function, allowing simultaneous extraction of film thickness and optical constants across the measured spectral range. To account for surface and interface morphology, additional layers representing surface roughness (modeled via Bruggeman effective medium approximation) and an intermix layer at the substrate interface were included. Model parameters such as the oscillator amplitude (Amp1), bandgap (Eg), and UV pole amplitude were refined through iterative fitting to minimize the mean square error (MSE) between measured and simulated $\Psi$ and $\Delta$ spectra. Multi-Sample Analysis (MSA) was employed to ensure consistency across related samples while capturing trends due to deposition condition changes. The resulting models exhibited good agreement with experimental data, yielding robust optical and structural parameters correlated with growth conditions. Representative fits of the ellipsometry measurements are presented in Figure S2.

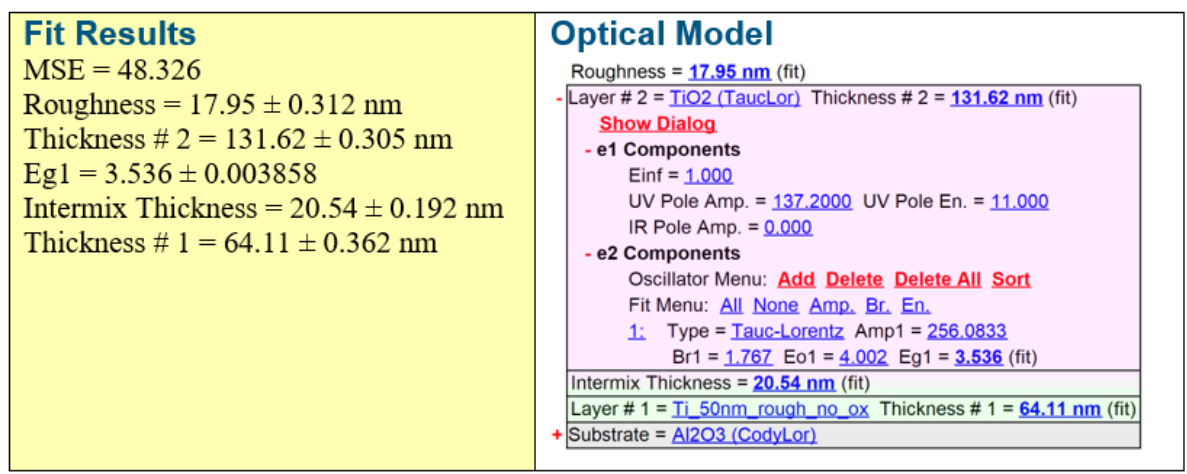

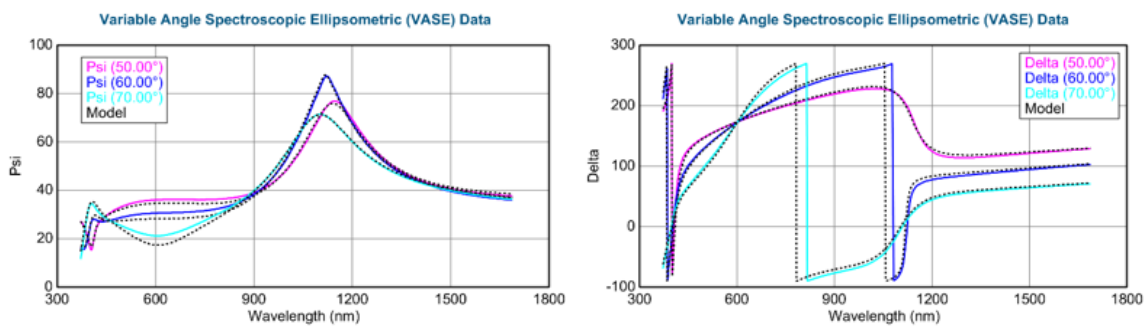

**Figure S2.** Representative ellipsometry data fitting and optical model used for $TiO_2$ thin film grown under 6.1 µbar and 8.3 Φ % $O_2$ synthesis conditions with HiPIMS.

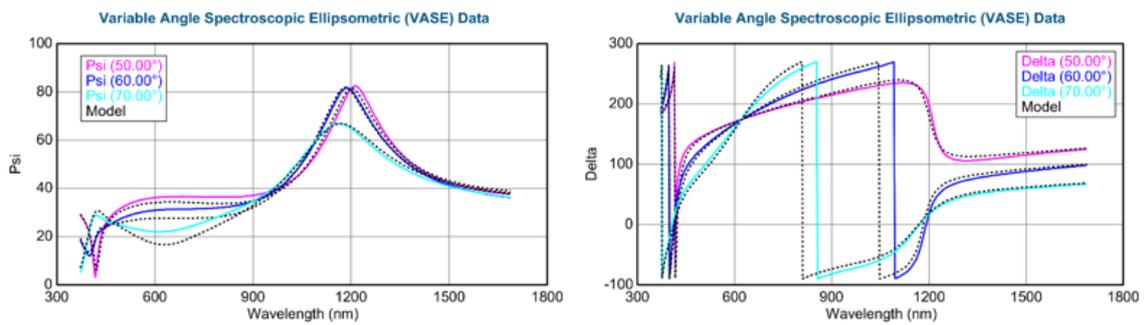

**Figure S3.** Representative ellipsometry data fitting and optical model used for $TiO_2$ thin film grown under 4.1 µbar and 12.9 Φ % $O_2$ synthesis conditions with HiPIMS.

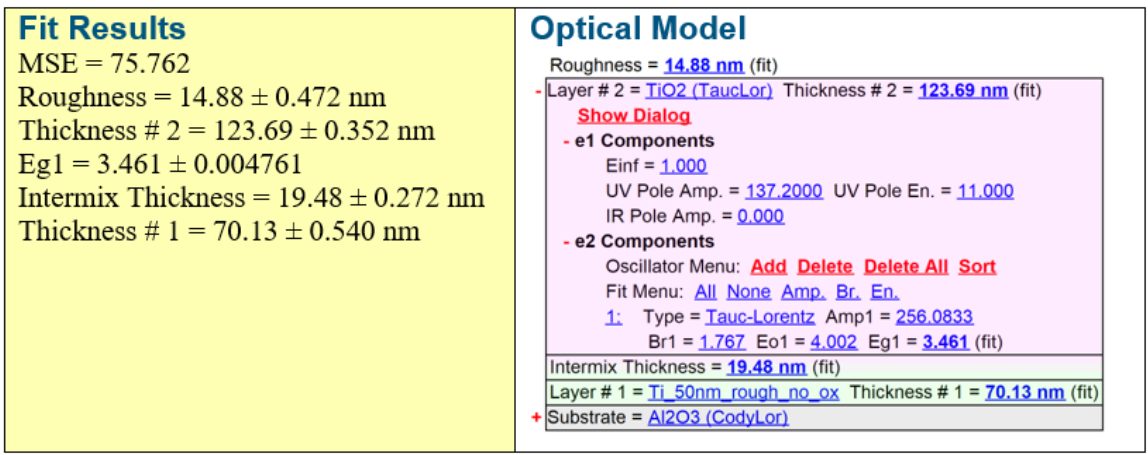

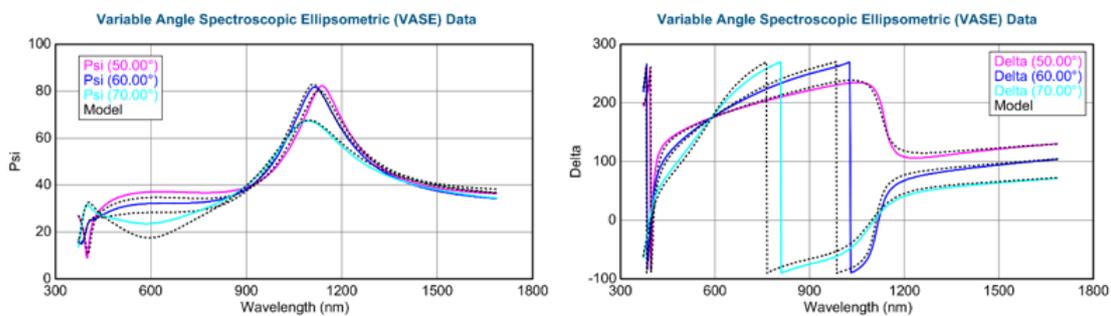

**Figure S4.** Representative ellipsometry data fitting and optical model used for TiO$_2$ thin film grown under 4.1 μbar and 31.4 Φ % O₂ synthesis conditions with HiPIMS.

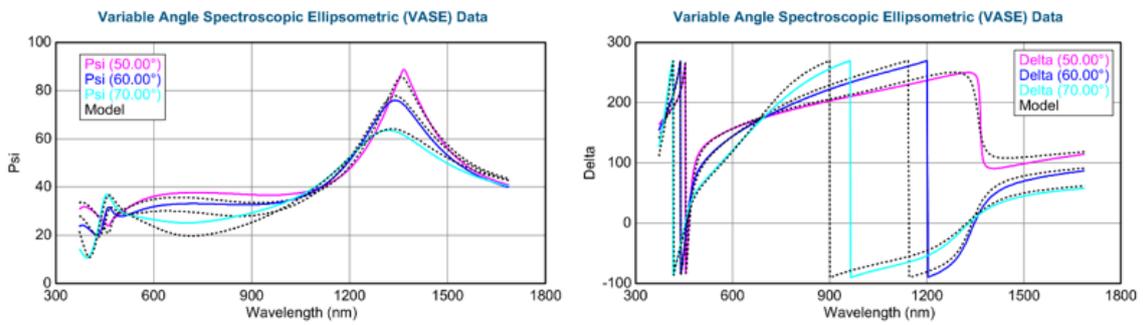

**Figure S5.** Representative ellipsometry data fitting and optical model used for TiO$_2$ thin film grown under 4.1 μbar and 53.3 Φ % O$_2$ synthesis conditions with HiPIMS.

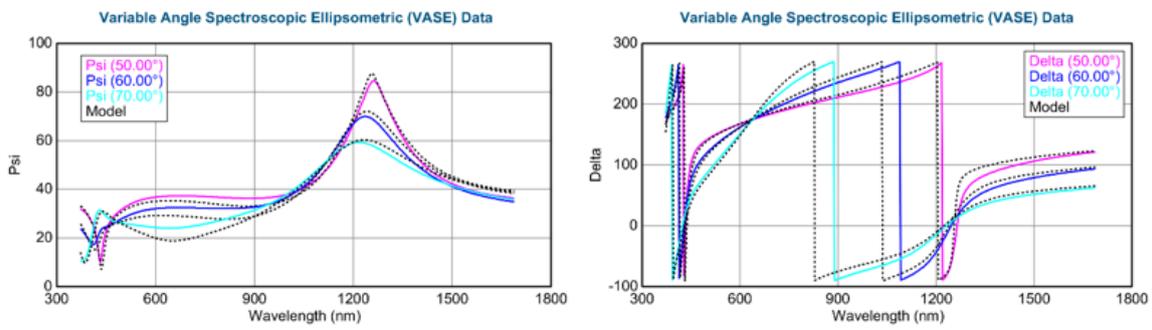

**Figure S6.** Representative ellipsometry data fitting and optical model used for TiO$_2$ thin film grown under 2.1 μbar and 15 Φ % O$_2$ synthesis conditions with HiPIMS.